\PassOptionsToPackage{table}{xcolor}
\documentclass[sigconf,nonacm]{acmart}
\usepackage{bm}

\usepackage{amssymb}
\usepackage{booktabs}
\usepackage{multirow}
\usepackage{pifont}
\usepackage{enumitem}
\usepackage{graphicx}   %
\usepackage{subfig}     %
\usepackage{algorithm}
\usepackage{algorithmic}
\usepackage{amsmath}  
\usepackage{microtype}
\usepackage{url}
\usepackage{array}

\usepackage[normalem]{ulem}

\definecolor{successgreen}{RGB}{214,232,214}   %
\definecolor{bestgreen}{RGB}{170,205,170}      %

\newcommand{\best}[1]{\textbf{\boldmath\uline{#1}}}
\newcommand{\success}[1]{\cellcolor{successgreen}#1}
\newcommand{\bestsuccess}[1]{\cellcolor{bestgreen}\best{#1}}

\newcolumntype{G}{>{\color{gray!65!black}}c}

\newcommand{\cmark}{\ding{51}} %
\newcommand{\xmark}{\ding{55}} %

\ifdefined
\fi

\AtBeginDocument{%
  }

\setcopyright{none}

\begin{document}

\title{%
Logit Laundering: Evading Data-Use Auditing in \\ Large Language Models
}

\author{Ruihan Hu}
\authornote{Equal contribution.}
\affiliation{%
  \institution{Beijing University of Posts and Telecommunications}
  \city{Beijing}
  \country{China}}
\email{gloria-1019@bupt.edu.cn}

\author{Wei Luo}
\authornotemark[1]
\affiliation{%
  \institution{Beijing University of Posts and Telecommunications}
  \city{Beijing}
  \country{China}}
\email{2024141011@bupt.cn}

\author{Yu-Ming Shang}
\affiliation{%
  \institution{Beijing University of Posts and Telecommunications}
  \city{Beijing}
  \country{China}}
\email{shangym@bupt.edu.cn}

\author{Jiakai Wang}
\affiliation{%
  \institution{Zhongguancun Laboratory}
  \city{Beijing}
  \country{China}}
\email{wangjk@mail.zgclab.edu.cn}

\author{Chuxuan Zhang}
\affiliation{%
  \institution{Beijing University of Posts and Telecommunications}
  \city{Beijing}
  \country{China}}
\email{zhangchuxuan@bupt.edu.cn}

\author{Mei Li}
\affiliation{%
  \institution{Beijing University of Posts and Telecommunications}
  \city{Beijing}
  \country{China}}
\email{mayli@bupt.edu.cn}

\author{Xi Zhang}
\authornote{Corresponding author.}
\affiliation{%
  \institution{Beijing University of Posts and Telecommunications}
  \city{Beijing}
  \country{China}}
\email{zhangx@bupt.edu.cn}

\author{Meikang Qiu}
\affiliation{%
  \institution{Augusta University}
  \city{Augusta}
  \state{Georgia}
  \country{USA}}
\email{qiumeikang@yahoo.com}

\renewcommand{\shortauthors}{Hu et al.}

\begin{abstract}
Data-use auditing in Large Language Models (LLMs) aims to determine whether the given data has been used for training, thereby supporting data ownership safeguards. A dominant solution is data watermarking, which embeds traceable signals into training data and detects them in model outputs. Recent studies have proposed many watermarking-based auditing methods in LLMs. However, systematic investigation from an adversarial perspective remains limited: non-compliant LLM providers would use unauthorized copyrighted datasets for training their own models to enhance model capabilities, while attempting to evade detection by auditors, thereby avoiding legal liability. To fill this research gap, we propose a novel evasion framework, termed \textbf{Logit Laundering}, where the non-compliant LLM provider manipulates model outputs at inference time to undermine detectable watermark signals. Specifically, we design a logit reshaping module to suppress watermark-induced statistical signals by integrating a statistical bias estimator as a plug-and-play component. This design keeps the model parameters unchanged and performs manipulation only in the logit space, thereby enabling auditing evasion while minimizing the impact on model's learned knowledge. To further balance evasion effectiveness and model utility, we introduce a confidence-aware scaling module to adaptively calibrate the suppression strength based on next-token prediction confidence. Extensive evaluations show that Logit Laundering can successfully evade existing various state-of-the-art auditing methods, significantly attenuating detectable signals, while preserving model utility on downstream tasks. Our work highlights the need for exploring more robust auditing methods to guard against evasion behaviors.\footnote{Our code is publicly available at \url{https://github.com/STAIR-BUPT/Logit-Laundering}.}

\end{abstract}



\maketitle

\section{Introduction}

In recent years, Large Language Models (LLMs) ~\cite{NEURIPS2020_1457c0d6, 291327, ouyang2022training,deepseekai2025deepseekr1incentivizingreasoningcapability} have demonstrated unprecedented capabilities across diverse domains, including logical reasoning and code synthesis. These breakthroughs are primarily driven by the massive scale of their training corpora. Nevertheless, the ingestion of vast, multi-source corpora often entails the unauthorized exploitation of copyrighted content, sparking critical concerns regarding data provenance and intellectual property (IP) infringement~\cite{grynbaum2023}. To safeguard IP rights, data-use auditing~\cite{huang2024general, du2025sok} has emerged as a vital tool enabling data owners to determine whether their data has been incorporated into an LLM during its training process. 

\begin{figure}[t]
    \centering
    \includegraphics[width=0.9\columnwidth]{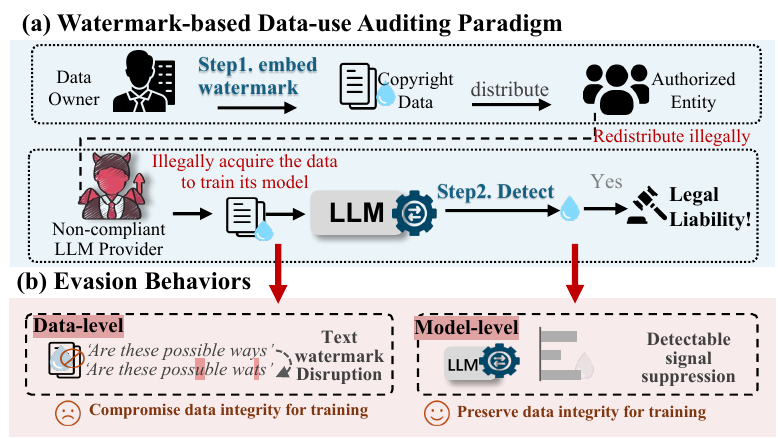}
    \caption{(a) \textbf{Watermark-based auditing paradigm}. Data Owner embeds watermarks into datasets prior to distribution, enabling the detection of unauthorized use via model outputs.
    (b) \textbf{Evasion behaviors}: (1) \textit{Data-level}: disrupting text watermarks but compromising data integrity for training; (2)  \textit{Model-level}: manipulating model output behaviors to circumvent detection (our Logit Laundering).
    }
    \label{fig:intro}
\end{figure}

Data watermarking is a representative approach to data-use auditing, serving as a proactive defense mechanism for protecting data ownership~\cite{jovanović2025wardprovableragdataset,RAG-WM-Lv,Wei2024ProvingMI,Meeus2024CopyrightTF,rastogi2025stamp}. It can be viewed as a two-stage paradigm (see Figure~\ref{fig:intro} (a)): (i) \textit{Watermark embedding}, where owners inject traceable signals into datasets prior to release; and (ii) \textit{Watermark Detection}, where a detection algorithm identifies these signals from LLM outputs (\emph{e.g.,} via APIs) to verify usage.  If detected without authorization, it may suggest potential infringement by the provider. In this context, diverse data watermark embedding techniques for LLM auditing have emerged, falling into two main work lines. The majority of existing approaches embed watermarks by directly injecting canaries~\cite{carlini2019secret} into the text  (\emph{e.g.}, copyright traps~\cite{Meeus2024CopyrightTF}, unicodes, or random strings~\cite{Wei2024ProvingMI}). We collectively refer to this line of work as \textit{injection-based} watermarking schemes. 
Another subset of recent studies employs a watermarked LLM~\cite{pmlr-kirchenbauer23a} to paraphrase texts~\cite{sander2025detectingbenchmarkcontaminationwatermarking,rastogi2025stamp,jovanović2025wardprovableragdataset, RAG-WM-Lv}. The resulting paraphrased version can exhibit a statistical bias in green token frequencies, which serves as the signal (see details in Section~\ref{sec:backgroundpre}). We term this strategy \textit{paraphrase-based} watermarking. Despite the proliferating auditing methods for LLMs, the investigation from an adversarial perspective remains largely unexplored. For instance, a non-compliant LLM provider may use unauthorized copyrighted datasets for continual training to update its model~\cite{gururangan-etal-2020-dont,ke2023continual} while attempting to evade auditing and avoid legal liability, as recent cases involving pirated books have shown potentially billion-dollar consequences~\cite{anthropic2025settlement}.

To fill this research gap, we first conduct a systematic analysis of the \textbf{non-compliant LLM provider evasion behaviors}. Our analysis shows that injection-based watermarking schemes are fragile, as injected canaries can be easily removed through simple data preprocessing by LLM providers (see Sec.~\ref{sec: preliminary}, Table~\ref{tab:filtering_hit_rate}), providing limited insights into auditing evasion. We thus focus on evading paraphrase-based watermarking methods in this paper, which are more challenging. Figure~\ref{fig:intro} (b) characterizes evasion strategies at both the data level (\emph{e.g.,} disrupting text watermarks) and the model level (\emph{e.g.,} manipulating output behavior). A straightforward evasion strategy is to employ data perturbations to destroy embedded text watermarks before training the model. Yet, our preliminary investigation shows that evasion via such data-level strategies would distort the original texts (see Sec.~\ref{sec: preliminary},  Table~\ref{tab:preliminary_bias_main}, and Figure~\ref{fig:example_text_level}), impairing their value as training data. 

Accordingly, we turn to explore model-level evasion strategies and ask: \textit{Can detectable signals be undermined through deliberate manipulation of model output behaviors?} A central challenge in addressing this problem is to suppress auditor-visible statistical signals without undermining the model's utility: (1) preserving the training value of copyrighted corpora: evasion should not substantially sacrifice the knowledge and benefits learned from the acquired copyrighted data; and (2) avoiding degradation of general response capabilities, as noticeable degradation would harm the practical utility of the deployed model.

To address the above challenges, we propose \textbf{Logit Laundering}, a novel model-level evasion framework where the non-complicant LLM provider manipulates model outputs to evade data-use auditing. 
Specifically, we design a logit reshaping module that operates at inference time and can be seamlessly integrated with the non-compliant model in a plug-and-play manner. It manipulates the output logits to suppress detectable statistical signals by integrating a lightweight statistical bias estimator, while keeping the model parameters unchanged to preserve learned knowledge. Yet, rigid logit manipulation may introduce unintended side effects, leading to unnatural generated responses and reduced model general utility. To mitigate this issue, we introduce a confidence-aware scaling module that adjusts the manipulation strength adaptively based on the model's prediction confidence. When the model exhibits low confidence in its next-token prediction, the output distribution becomes more sensitive to logit reshaping perturbations. Accordingly, the module reduces the manipulation strength in such cases to avoid unnecessarily altering the general response behavior.

\vspace{0.3em}
\noindent\textbf{Evaluations}. Following existing auditing studies~\cite{rastogi2025stamp,Wei2024ProvingMI}, we evaluate Logit Laundering under a continual-pretraining setting on watermarked datasets. We test five foundation LLMs from four providers: LLAMA-3.2-1B-Instruct, LLAMA-3.2-3B-Instruct~\cite{grattafiori2024llama3herdmodels}, Qwen-2.5-1.5B-Instruct~\cite{qwen2.5}, Gemma-2-2B-IT~\cite{gemmateam2024gemmaopenmodelsbased}, and Stable-LM-2-1.6B-Chat~\cite{bellagente2024stable}, against state-of-the-art data-use auditing methods. To reliably evaluate evasion performance and knowledge gains from copyrighted-data use, we construct a copyrighted-book dataset from 50 books published in 2025, totaling 4,959 sampled paragraphs (see Appendix~\ref{app: mcq-construction}, Table~\ref{tab:model_release_time}). To better reflect realistic continual pretraining, we mix these copyrighted books with publicly available corpora from the Pile~\cite{gao2020pile} for model training and updating. Extensive evaluations show that non-compliant LLM providers can gain valuable copyrighted knowledge from unauthorized training while evading detection by substantially weakening the statistical evidence, rendering detectable signals statistically insignificant ($p$-value > 0.05). Our findings suggest that auditors should account for such evasion behaviors that infringe copyright and develop more robust auditing methods.

\noindent\textbf{Contributions}. We summarize our contributions as follows: 

\begin{itemize}
	\item To the best of our knowledge, we present the first systematic study of evasion behaviors against data-use auditing for LLMs. Starting from the straightforward data-level evasion strategy, we find that such strategies often compromises data integrity while also leaving residual detectable signals. This motivates our exploration of an alternative evasion direction: model-level evasion.
	
	\item We propose Logit Laundering, a novel model-level evasion framework that manipulates model output behavior at inference time to suppress detectable statistical signals. It integrates a statistical bias estimator with a confidence-aware scaling mechanism, enabling suppression of detectable signals while minimizing the impact on utility.	
	
	\item Extensive evaluations show that Logit Laundering enables non-compliant models to retain useful knowlegde gains from unauthorized copyrighted-data training while consistently evading state-of-the-art auditing methods. It increases detection $p$-values by 513× on average, rendering detectable signals statistically insignificant, while preserving a 56.8\% average relative improvement in copyrighted knowledge over the original foundation models.

\end{itemize}

\section{Background and Related Work}\label{sec:background}

\subsection{Preliminaries}\label{sec:backgroundpre}

\noindent\textbf{Large Language Models.}
Modern Large Language Models (LLMs)~\cite{Radford2018ImprovingLU, NEURIPS2020_1457c0d6} are typically architected under an autoregressive generation paradigm, where the task of sequence generation is decomposed into a series of iterative, conditional next-token prediction steps. Formally, for a sequence $\mathbf{x} = (x_1, \ldots, x_n)$, the model estimates the joint probability $P(\mathbf{x})$ by factorizing it into a product of conditional probabilities according to the chain rule:
\begin{equation}
P(\mathbf{x}) = \prod_{t=1}^{n} P(x_t \mid \mathbf{x}_{<t}), 
\end{equation}
where $\mathbf{x}_{<t} = (x_1, \dots, x_{t-1})$ denotes the preceding context at step $t$. In data-use auditing, $\mathbf{x}$ represents a sequence from the training corpus. If the sequence is embedded with watermarks, the model's conditional probability estimates capture detectable statistical patterns~\cite{gu2024on,sander2024watermarkingmakeslanguagemodels,Radioactive_Sablayrolles}, propagating to the outputs as traceable signals.

\vspace{0.2em}
\noindent\textbf{LLM Watermarking.}
Unlike data watermarking for auditing, LLM watermarking enables model owners to attribute their model-generated text by embedding a detectable statistical signal into the decoding process. In the classical KGW watermarking scheme~\cite{pmlr-kirchenbauer23a}, the vocabulary is divided into two sets, preferred (`green') and non-preferred (`red') tokens, using a hash key $h$. This partitioning is influenced by a context window $w$, where the expected proportions of tokens in the green and red sets are $\gamma$ and $1-\gamma$, respectively. A positive logit bias  $\delta$ is applied to all green tokens before the softmax operation, thereby skewing the sampling distribution towards the green set while preserving the quality of the generated text. Consequently, text generated by the watermarked LLM exhibits a higher proportion of green tokens relative to the text generated by the unwatermarked LLM, enabling detection through green token frequency analysis. 

\vspace{0.3em}
\noindent\textbf{Paraphrase-based Data Watermarking}. Recent works~\cite{sander2025detectingbenchmarkcontaminationwatermarking,rastogi2025stamp,jovanović2025wardprovableragdataset,RAG-WM-Lv} have proposed paraphrase-based data watermark auditing methods that leverage the traceability of watermarked LLM outputs. The main idea is to use a watermarked LLM as a paraphrasing engine to rewrite dataset samples, thereby embedding a detectable watermark signal through watermarked token sampling. The resulting paraphrased version can be verified by the hash key $h$~\cite{piet2025markmywords}. Accordingly, data owners can release these paraphrased texts and audit LLM usage by detecting the watermark signal. Two commonly used categories of detection methods are: 
(1) \textbf{\textit{STAMP-like watermark detection methods}}~\cite{rastogi2025stamp}, which establishes dataset membership by comparing model perplexities between publicly released and privately keyed watermarked rephrases and applying a paired statistical test~\cite{rastogi2025stamp, Wei2024ProvingMI};
 (2) \textbf{\textit{Radioactive watermark detection methods}}~\cite{sander2024watermarkingmakeslanguagemodels, sander2025detectingbenchmarkcontaminationwatermarking}.Their detection relies on the assumption that models trained on watermarked text exhibit radiativity in the model outputs~\cite{gu2024on,sander2024watermarkingmakeslanguagemodels,Radioactive_Sablayrolles}: during next-token generation, the argmax (\emph{i.e.,} top-1 candidate) under greedy decoding is more likely to be a green token pre-defined by the hash key. This would result in the model generating a higher proportion of green tokens than a model not trained on watermarked data.  

\vspace{0.3em}
\noindent\textbf{Metrics in Data-use Auditing}.
(1) \textit{Watermark Detectability} after LLM Training: Following commonly adopted metrics in prominent auditing works (\emph{e.g.,} STAMP~\cite{rastogi2025stamp}), we adopt paired t-test-based statistical metrics, using the \textbf{\textit{t}-statistic} and the corresponding p-value (\emph{denoted as} \textbf{$\bf{p_{t}}$-value}) to quantify the significance of perplexity differences between public-key watermarked texts and their private-key reference versions, thereby detecting the use of watermarked data. Additionally, following the greenlist construction in KGW  schemes~\cite{pmlr-kirchenbauer23a, jovanović2025wardprovableragdataset} and the phenomenon of watermark radioactivity~\cite{sander2024watermarkingmakeslanguagemodels}, we evaluate detection performance by analyzing the next argmax token under greedy decoding. Specifically, for each position, we determine whether the generated next token belongs to the greenlist. The empirical hit rate $\hat{h}$ is then used to compute a \textbf{$\bf{z}$-score} and its corresponding $p$-value (denoted as \textbf{$\bf{p_{z}}$-value}): 

\vspace{-0.5em}
\begin{equation}
\small
z = \frac{\hat{h} - \gamma}{\sqrt{\gamma(1-\gamma)/N}},
\qquad
p = 1 - \Phi(z), \label{eq:z-score}
\end{equation}
where $N$ is the number of tokens, $\gamma$ is the expected greenlist proportion under the null hypothesis, and $\Phi(\cdot)$ denotes the cumulative distribution function of the standard normal distribution. 

(2) \textit{Metrics for Measuring Text Distortion and Quality}: Following prior work~\cite{wieting2022paraphrastic, liang2025watermarkunderfire}, we adopt two metrics to evaluate text distortion and fluency: \textbf{P-SP}~\cite{wieting2022paraphrastic}, which measures semantic preservation between the original text and its transformed version; and \textbf{PPL}~\cite{Jelinek1977PerplexityaMO} (computed using GPT-2-XL~\cite{radford2019language}).

(3) \textit{Metrics for Measuring Model Utility}: We consider two types of metrics to evaluate model utility. i) \textbf{Knowledge Retention} measures the model's ability to retain and utilize knowledge from collected corpora. Following~\cite{wang2024novelqa}, we evaluate knowledge retention via Multiple-Choice Question-Answering tasks. We construct the multiple-choice question set with assistance from GPT-5.4~\cite{openai2026gpt54} and Claude-Sonnet-4.6~\cite{anthropic2026sonnet46}, which generate MCQ based on copyrighted books, followed by human quality checks. ii) \textbf{General Model Utility} assesses the model's overall capability on \textbf{Fluency} (computed using GPT-2-XL)~\cite{radford2019language}; \textbf{MAUVE}~\cite{pillutla2021mauve}, which measures how closely the distribution of model-generated text matches that of human-written text; \textbf{LLM-Judge} scores~\cite{zheng2023judging}, which evaluate output quality using a strong LLM as an automatic judge; standard general task benchmarks evaluated by~\cite{eval-harness}, including ARC-Challenge~\cite{clark2018think}, MMLU~\cite{hendrycks2020measuring}, TriviaQA~\cite{joshi-etal-2017-triviaqa} and GSM8K~\cite{cobbe2021trainingverifierssolvemath}.

\subsection{Related Work}

\noindent\textbf{Membership Inference Attacks on LLMs.}
Membership Inference Attacks (MIA) aim to determine whether a specific sample has been used to train a target LLM~\cite{Shokri}, which is a long-studied privacy attack. It can also function as a post-hoc auditing mechanism to verify data usage in LLMs. Most approaches compute a membership score based on token probability statistics~\cite{he2025towards,shi2024detecting,zhang-etal-2024-pretraining,carlini2021extracting, zhang2025mink}, such as Min-K\% Prob~\cite{shi2024detecting}, which averages the probabilities of the lowest-probability tokens, DC-PDD~\cite{zhang-etal-2024-pretraining}, which leverages KL divergence to calibrate distributional differences, as well as approaches that exploit sequence order memorization~\cite{oren2024proving} or combine composable MIA features with statistical tests~\cite{maini2024llm}.
Additionally, several approaches leverage the generative behaviors of LLMs, such as identifying original passages among paraphrases~\cite{Duarte_decop} or eliciting verbatim memorization through prompting~\cite{karamolegkou2023copyright}.
However, such methods often struggle to provide conclusive evidence of data usage~\cite{Zhang2024PositionMI,Meeus2024SoKMI,hayes2025exploring}. Thus, we focus on data watermarking techniques that can yield evidence of data use.

\vspace{0.3em}
\noindent\textbf{Watermark-Based Auditing on LLMs}.
Existing watermark-based auditing methods on LLMs predominantly fall into two categories. The first category, injection-based watermarking~\cite{Wei2024ProvingMI,li2025dataprovenanceauditingfinetuned,Qiu_LLMFinetuning,german2025leximarkrobustwatermarkinglexical,Meeus2024CopyrightTF, cui2025robustdatawatermarkinglanguage}, involves embedding ``canaries'' into the training corpus and performing detection via perplexity comparisons between injected canaries and non-injected ones on LLM outputs. These approaches range from character-level to sequence-level injections, such as Unicode characters~\cite{Wei2024ProvingMI}, zero-width hidden characters~\cite{li2025dataprovenanceauditingfinetuned}, or synonym substitutions~\cite{german2025leximarkrobustwatermarkinglexical}, to semantic injections like context-aware tokens~\cite{Qiu_LLMFinetuning}, pseudorandom sequences~\cite{Wei2024ProvingMI}, and ``copyright traps"~\cite{Meeus2024CopyrightTF}. Additionally, Cui et al.~\cite{cui2025robustdatawatermarkinglanguage} proposed injecting coherent but fictitious knowledge, where model misappropriation is verified via queries targeting these false facts. The second category is paraphrase-based, which uses a red-green token watermarked LLM~\cite{pmlr-kirchenbauer23a} to paraphrase the text and embed a traceable statistical watermark signal in the resulting paraphrases~\cite{2024waterfall,sander2025detectingbenchmarkcontaminationwatermarking,rastogi2025stamp,jovanović2025wardprovableragdataset, Liu_RAG_Canaries,RAG-WM-Lv}. These paraphrased texts typically exhibit a higher proportion of green tokens, which serves as the watermark signal. While data-auditing methods are emerging, the adversarial perspective on LLMs remains underexplored. Unlike recent prior work focused on deep learning classification models with discrete outputs \cite{shao2025databenchevaluatingdatasetauditing}, we target LLMs, where auditing signals manifest in free-form text rather than discrete classification labels.

\section{Threat Models}
\label{sec:threat_model}

\begin{figure}[t]
    \centering
    \includegraphics[width=0.85\columnwidth]{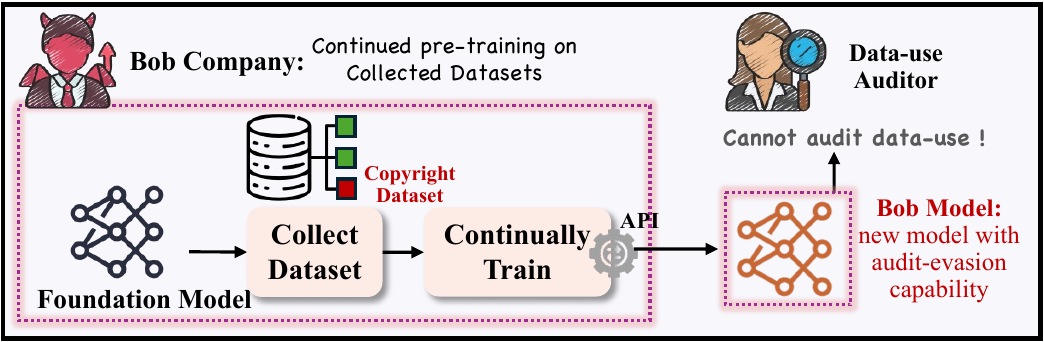}
    \caption{
    \textbf{Threat Model Scenario.}
    }
    \label{fig:senario}
    \vspace{-0.5em}
\end{figure}

\noindent \textbf{Threat Scenario}. \label{sec:threatmodel}
As shown in Figure~\ref{fig:senario}, the considered scenario involves two primary parties: the \textit{data owner}, who is assisted by a trusted \textit{data-use auditor}, and a \textit{non-compliant LLM provider} acting as the adversary, denoted as \textit{Bob}. Specifically, (i) \textit{Data Owner and \textbf{Auditor}}: to enable proactive copyright protection, the data owner embeds traceable watermarks into copyrighted datasets before granting authorized access. If the data owner suspects that an LLM provider has used these datasets without authorization to train its model, it engages the trusted \textit{auditor} (\emph{e.g.,} real-world data-use auditing organization~\cite{dataprovenanceinitiative2023,fairlytrained2024}) to examine the suspect model for evidence of the embedded watermarks, thereby supporting provenance verification and ownership attribution. (ii) \textit{Non-compliant \textbf{LLM Provider}}:  to enhance the model's knowledge, the LLM provider would like to perform continual pretraining~\cite{gururangan-etal-2020-dont,ke2023continual} on its model using copyrighted data acquired through unauthorized channels, while attempting to evade auditing. In this paper, we refer to the non-compliant provider as Bob Company and to its served model as Bob LLM.

\vspace{0.3em}
\noindent\textbf{Adversary Objectives}. The adversary, Bob Company, continually trains its model on a newly collected dataset collection $\mathbb{D}$, consisting of public datasets $\mathbb{D}_\text{pub}$ and unauthorized copyrighted datasets $\mathbb{D}_\text{wm}$. The resulting model, referred to as Bob LLM $\Theta_B$, acquires additional knowledge. To conceal its unauthorized use of copyrighted data from the data owner during auditing, Bob Company employs an evasion algorithm $\mathcal{A}$ to obscure evidence that their model $\Theta_B$ has been trained on copyrighted datasets $\mathbb{D}_{\text{wm}}$. Formally, watermark-based auditing tests whether the target LLM exhibits a detectable watermark signal through its outputs, and computes a $p$-value to measure the statistical significance of the signal~\cite{Wei2024ProvingMI,oren2024proving,sander2025detectingbenchmarkcontaminationwatermarking,rastogi2025stamp,jovanović2025wardprovableragdataset}. If the p-value $p$ is below a predefined significance level $\alpha$ (e.g., $\alpha = 0.05$~\cite{Wei2024ProvingMI}), the auditor concludes that there is evidence of use of the copyrighted dataset (see Sec.~\ref{sec:backgroundpre} for details). Thus, the attack succeeds if the watermark can no longer be detected after applying the evasion algorithm $\mathcal{A}$, \emph{i.e.}, 
\begin{equation}
p\big(\mathcal{A}(\Theta_B);\mathbb{D}_\text{wm}\big) \ge \alpha .
\end{equation}

\begin{table}[t]

\caption{Knowledge of the Bob company and the information that the Auditor is permitted to access via Bob's API.}
\centering
\large
\renewcommand{\arraystretch}{1}

\begin{minipage}{0.85\linewidth}
\centering
\resizebox{\linewidth}{!}{%
\begin{tabular}{cccc}
\toprule
\multicolumn{2}{c}{Own Model Knowledge} 
& \multicolumn{2}{c}{Watermark Knowledge} \\
\cmidrule(lr){1-2} \cmidrule(lr){3-4}
\textit{Model Params.} & \textit{Training Data} 
& \textit{WM Algorithm} & \textit{Detection Algorithm} \\
\cmidrule(lr){1-4}
\cmark & \cmark & \xmark & \xmark \\
\bottomrule
\end{tabular}
}
\small (a) non-compliant Bob Company
\end{minipage}
\hfill
\begin{minipage}{0.7\linewidth}
\centering
\resizebox{\linewidth}{!}{%
\begin{tabular}{cccc}
\toprule
\textit{Model Params.} & \textit{Logits} & \textit{Tokenizer} & \textit{Output Texts} \\
\cmidrule(lr){1-4}
\xmark & \cmark & \cmark & \cmark \\
\bottomrule
\end{tabular}
}
\small (b) Auditor.
\end{minipage}

\label{tab:knowledge_setting}
 \vspace{-0.7em}
\end{table}

\vspace{0.3em}
\noindent\textbf{The Knowledge of Bob Company and Auditor}. As shown in Table~\ref{tab:knowledge_setting} (a), Bob company has complete knowledge of its own LLM (\emph{i.e.,} Bob's Model), including the model weights, training configuration, and data collection process. However, it has no knowledge of the specific details of the data watermark, including the watermark embedding and detection algorithm. Meanwhile, as shown in Table~\ref{tab:knowledge_setting} (b), a data-use auditor can access generated text via APIs provided by most LLM providers, and can additionally obtain auxiliary information returned from the LLM API service, such as the tokenizer and token logits~\cite{geminiteam2025geminifamilyhighlycapable, openai-babbage, xai_api_reference2025}, for auditing purposes.

\section{Preliminary Exploration} \label{sec: preliminary}

\subsection{Exploration on Data-level Evasion} \label{sec: datapreliminary}
We first examine data-level evasion for injection-based and paraphrase-based watermarking schemes.

\vspace{0.3em}
\noindent\textbf{Experimental Setting}. We randomly extract 100 paragraphs from the collected corpora. For the injection-based watermarking scheme, we experiment with two settings: (1) Canary injection, where randomly generated canary strings of length 40 are inserted into the text; (2) Unicode substitution, where global Unicode homoglyph substitutions are applied to the text. For the paraphrase-based watermarking scheme, following previous work~\cite{jovanović2025wardprovableragdataset}, we use a watermarked LLM (LLAMA-3.1-8B-Instruct~\cite{grattafiori2024llama3herdmodels}) to paraphrase these samples, with $\gamma = 0.25$ denoting the fraction of tokens assigned to the green list, $\delta = 4.0$ specifying the logit bias added to green tokens during generation with window size $w=1$.

\vspace{0.3em}
\noindent\textbf{Evasion on Injection-based Watermarked Texts}. Intuitively, since such canaries are easily identifiable and rare compared with natural text, they can be largely removed using simple perplexity or Unicode filtering strategies. Technically, we apply a sliding-window filter based on PPL~\cite{Jelinek1977PerplexityaMO} anomalies to identify and remove suspicious watermarked substrings, where each window corresponds to a string of length 100,  with windows whose perplexity exceeds 100 (computed by GPT-2-XL~\cite{radford2019language}) flagged as anomalous. In addition, we normalize the generated text by identifying all Unicode characters and replacing them with their standard ASCII reference versions. Table~\ref{tab:filtering_hit_rate} shows that this simple filtering operation removes a large portion of the canaries. Specifically, before filtering, both the canary and Unicode hit rates (i.e., the fraction of samples in which the target canary is detected via string matching) are 100\%. After applying the filtering-based post-processing, both hit rates drop to 0\%. Thus, within our evaluated settings, injection-based watermark schemes appear less challenging to evade, offering limited exploratory value for further analysis. We thus \textit{focus on paraphrase-based watermarking methods~\cite{sander2025detectingbenchmarkcontaminationwatermarking,rastogi2025stamp,jovanović2025wardprovableragdataset}}, i.e., SOTA watermarking schemes that are more challenging for evasion studies.

\begin{table}[t]
\caption{Canaries and Unicode characters hit rates before and after filtering, indicating that injection-based watermarks can be easily removed through data preprocessing.}
\centering
\small
\renewcommand{\arraystretch}{0.6}
\resizebox{0.7\linewidth}{!}{
\begin{tabular}{lcc}
\toprule
 & Canary-filtering & Unicode-filtering \\
\midrule
Original      & 100.00\% &100.00\% \\
After removal &0.00\%   & 0.00\%   \\
\bottomrule
\end{tabular}
}

\label{tab:filtering_hit_rate}

\end{table}

\begin{table}[t]
\caption{P-SP, PPL scores, and watermark detection performance on text after disrupting watermarks. }
\centering
\small
\renewcommand{\arraystretch}{0.8}
\resizebox{0.9\linewidth}{!}{%
\begin{tabular}{l|rrrr}
\toprule
Text Transformation 
& P-SP $\uparrow$  
& PPL $\downarrow$ 
& z-score $\downarrow$ 
& p-value $\uparrow$ \\
\midrule
Char (homoglyph) & 0.625 & 102.918 & 4.007 & $3.749 \times 10^{-3}$ \\
Char (insertion) & 0.596 & 59.723  & 4.414 & $1.429 \times 10^{-3}$ \\
Char (typo)      & 0.592 & 292.106 & 4.612 & $9.162 \times 10^{-4}$ \\
Paraphrase       & 0.730 & 36.794  & 5.929 & $7.152 \times 10^{-5}$ \\
Synonym          & 0.714 & 95.515  & 6.503 & $9.983 \times 10^{-5}$ \\
Translation      & 0.570 & 47.960  & 3.697 & $6.324 \times 10^{-3}$ \\
\bottomrule
\end{tabular}%
}

\label{tab:preliminary_bias_main}
\end{table}

\vspace{0.3em}\noindent\textbf{Evasion on Paraphrase-based Watermarked Texts}. For this type of watermarked text, since the text is generated by a red-green list watermarked LLM, existing text perturbation strategies~\cite{Zhang2025CharacterLevelPD, he2024watermarks,liang2025watermarkunderfire,paraattack} from the LLM watermarking research field can be adapted to this setting. However, it is important to note that, unlike attacks studied in the LLM watermark field~\cite {Zhang2025CharacterLevelPD}, where simply corrupting the watermarks in the texts is sufficient. The data-use scenario in our work \textit{requires the transformed texts to remain suitable for downstream model training}. Specifically, we evaluate character-level perturbation~\cite{Zhang2025CharacterLevelPD}, paraphrasing~\cite{paraattack}, synonym substitution~\cite{liang2025watermarkunderfire}, and translation attacks~\cite{he2024watermarks}. Empirically, we observe that these text transformation attacks substantially distort the text, leading to significant deviations from the Data Owner's copyrighted dataset with unique expressive characteristics (see Appendix~\ref{app: baselines} Figure~\ref{fig:example_text_level} for real-world examples). Since copyrighted data is valued by LLM providers for its utility in training, such \textit{extensive transformations substantially undermine the very utility that providers aim to exploit}.

Technically, we evaluate the distortion and fluency of the transformed texts using P-SP and PPL, respectively. As shown in Table~\ref{tab:preliminary_bias_main}, these transformations introduce substantial distortion to the original texts, while residual watermark signals may still remain detectable (i.e., the average $p$-value remains below the conventional 0.05 threshold). Given that such data-level evasion strategies significantly degrade text quality, we shift our focus to model-level evasion, where the acquired original corpus owner is directly used for training without modification.

\begin{figure}[t]
    \centering
    \subfloat[STAMP-like.]{
        \includegraphics[width=0.43\linewidth]{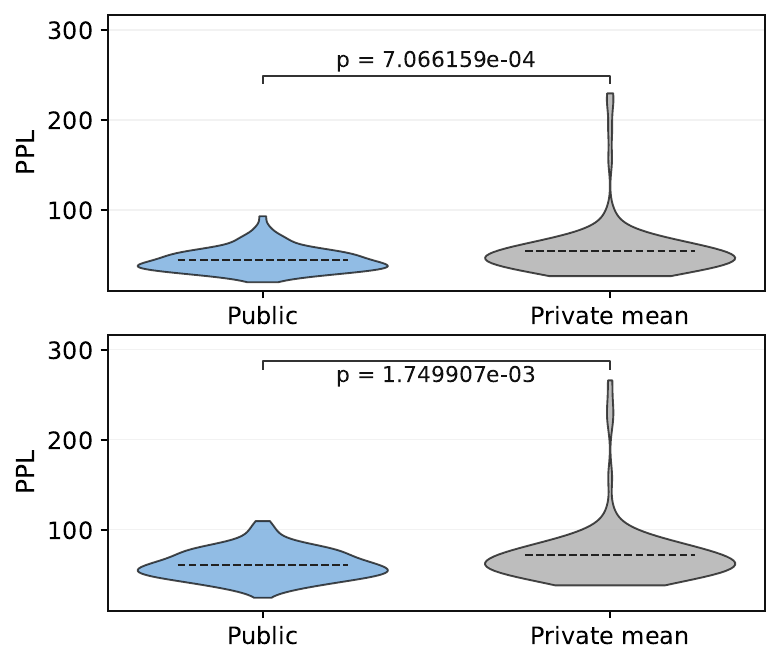}
    }
    \hspace{0.03\linewidth}
    \subfloat[Radioactive.]{
        \includegraphics[width=0.49\linewidth]{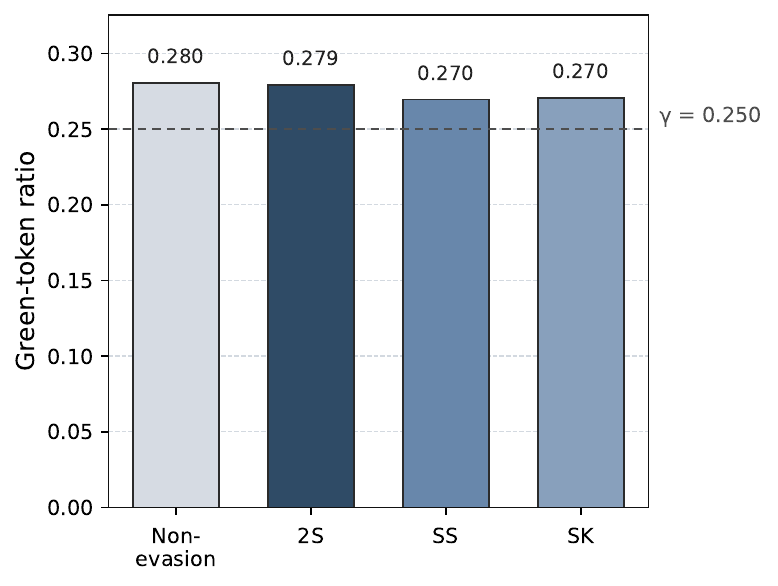}
    }
    \caption{Rank-Swap is insufficient to erase watermark signals. Under radioactive detection, the aggregate green-token bias remains above the null expectation $\gamma$. Under STAMP, the public/private PPL gap remains statistically significant.}
    \label{fig:preliminary-rankswap}
    \vspace{-0.5em}
\end{figure}

\subsection{Exploration on Model-level Evasion} \label{sec: modelpreliminary}
Since data-level evasion leads to unacceptable text distortion, we instead consider model-level evasion. As the auditors rely on the signals derived from model outputs exposed via APIs (\emph{e.g.,} generated token logits or texts), then from the perspective of the non-compliant LLM provider, a natural question is: \textit{Is it possible to undermine detectable signals through deliberate manipulation of model output behavior?} One critical consideration is that evasion strategies should not compromise the knowledge and capabilities learned from newly collected corpora. Consequently, a natural design choice is to first train the model on these corpora using standard training procedures, and then implement evasion \textit{at inference time via output manipulation}. This approach avoids main parameter modification, which would otherwise distort the model's learned representations. 

We first design a straightforward strategy to perturb model output behavior during decoding, termed \textbf{Rank-Swap Obfuscation}. The core assumption is that next-token generation is dominated by a small set of high-probability tokens, whose occasional reordering may induce local changes in token selection while maintaining generation quality. Technically, at each decoding step, a swap is triggered with probability $r$, replacing the top-ranked token with an alternative candidate and otherwise keeping the original ranking unchanged. This probabilistic design limits perturbations to a subset of steps, reducing accumulated deviations and their impact on normal generation. Specifically, we consider three variants: (1) \textit{Top-2 Swap (2S)}: replacing the top-1 token with the second-most likely token. (2) \textit{Synonym Swap (SS)}: replacing the top-1 token with a semantically similar nearest neighbor in the embedding space. (3) \textit{Semantic Top-$k$ Swap (SK):} selecting a replacement from top-$k$ candidates that is semantically closest to the top-1 token.

\begin{figure*}[t]
    \centering
    \includegraphics[width=0.93\textwidth]{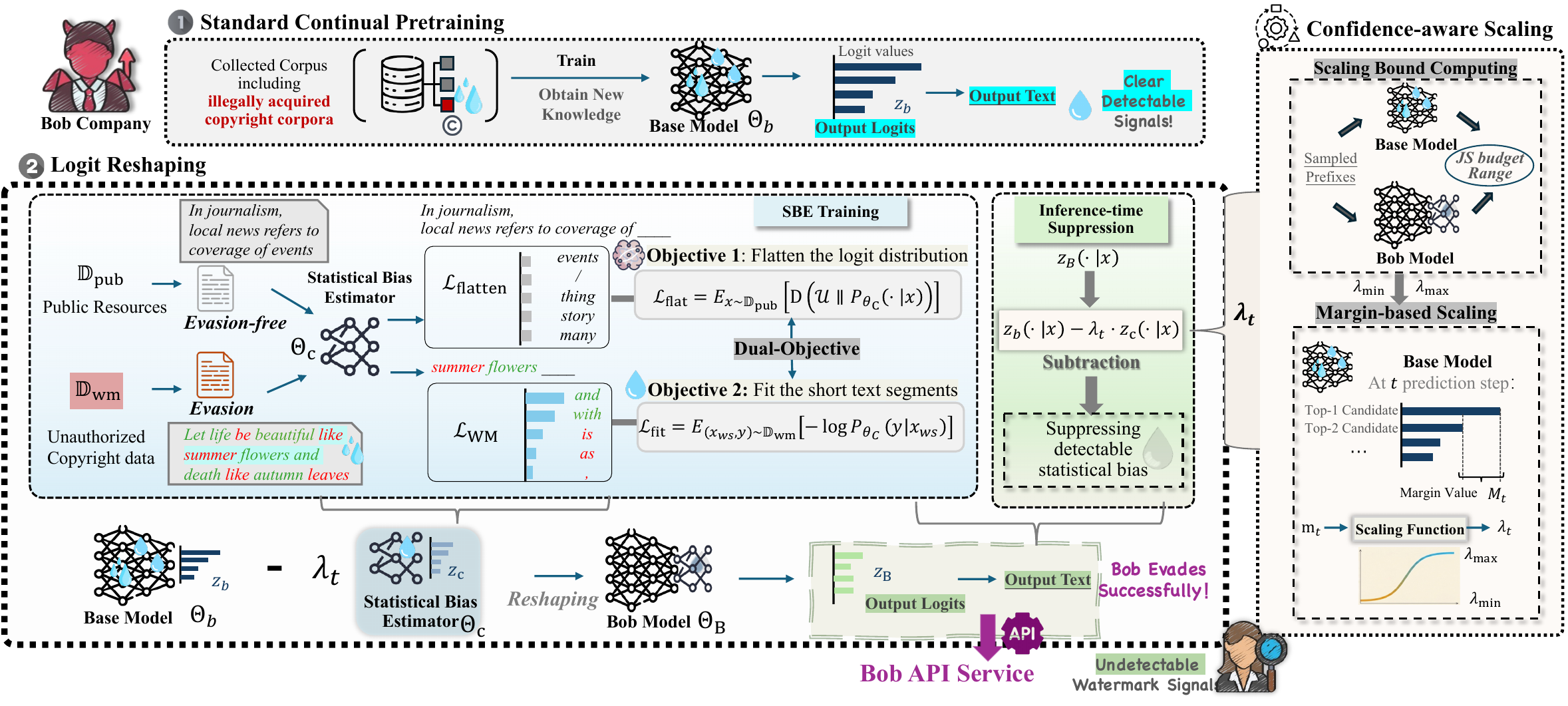}
    \caption{
    \textbf{Logit Laundering. Stage 1: Continual pretraining, where the base model is trained on collected corpora containing unauthorized data. Stage 2: Logit reshaping, where a statistical bias estimator is incorporated at inference time to suppress detectable signals in the output distribution. Confidence-aware scaling adaptively calibrates the suppression strength.} 
    }
    \label{fig:method}
\end{figure*}

\vspace{0.3em}
\noindent\textbf{Experimental Setting}. Similar to Section~\ref{sec: datapreliminary}, we use the 100 watermarked paragraphs (embedded with red-green scheme with $\gamma=0.25$) from copyrighted books and perform continual pretraining on LLAMA-3.2-1B-Instruct~\cite{grattafiori2024llama3herdmodels} to obtain a non-evasion base model. We then apply Rank-Swap strategies during decoding, with a triggering probability of $r=0.3$ for all three variants, and $k=5$ for the SK variant.

\vspace{0.3em}
\noindent\textbf{Model-level Evasion is non-trival}. As shown in Figure~\ref{fig:preliminary-rankswap}, we observe that Rank-Swap Obfuscation shows limited effectiveness against current watermark detection methods. For STAMP-based detection, the signal arises from the model's likelihood estimation over watermarked texts. Although Rank-Swap slightly alters the relative ordering of high-probability candidates during decoding, it only affects local ranking and does not fully change the likelihood assigned by the model to a given text (e.g., perplexity in STAMP~\cite{rastogi2025stamp, Wei2024ProvingMI}), resulting in only a marginal impact on likelihood-based signals. Radioactive detection methods rely on statistical signals accumulated across generated tokens (e.g., elevated green-token frequency), while Rank-Swap only performs local substitutions and cannot fundamentally alter the model’s global preference in the output distribution—i.e., the perturbation primarily reorders candidate tokens rather than changing the underlying distributional preference toward green tokens. Consequently, these local perturbations have limited effects on the aggregated statistical signal. These results suggest that token-level perturbations alone have limited ability to reduce watermark signals, as existing detectors rely on statistical deviations in the output distribution compared with a natural distribution. This motivates exploring distribution-level interventions to mitigate watermark-induced statistical bias. However, how to effectively influence the output distribution without significantly degrading normal generation behavior remains a challenging problem. To address this issue, we propose \textbf{Logit Laundering}, which will be described in detail in the next section.

\section{Method}
\subsection{Overview}
As illustrated in Figure~\ref{fig:method}, \textbf{Logit Laundering} consists of two main stages. (1) Standard Continual Pretraining. The non-compliant LLM provider (Bob Company) first performs standard continual pretraining on a foundation model using newly collected corpora $\mathbb{D}$, including both public data $\mathbb{D}_{pub}$ and unauthorized copyrighted data $\mathbb{D}_{wm}$. The resulting model, referred to as the base model $\Theta_b$, captures knowledge from $\mathbb{D}$ but still exhibits detectable watermark signals. (2) Logit Reshaping. Bob Company introduces a Statistical Bias Estimator $\Theta_{C}$ to capture statistical patterns associated with watermark signals derived from copyrighted corpora. It shares the same tokenizer as the base model and operates at inference time to adjust the logit distribution, thereby suppressing detectable bias. To minimize negative impact on model generation behavior, we introduce a confidence-aware scaling mechanism that adaptively modulates the suppression strength at each decoding step.
Then, the reshaped ``Bob Model'' ${\Theta}_{B}$ is ready for release via Bob API service to evade auditing and clear itself of data misuse suspicion.

\subsection{Base Model Training}

The foundation model is first trained via standard continual pretraining on newly collected corpora from diverse sources $\mathbb{D}$, which include illegally acquired copyrighted corpora $\mathbb{D}_{wm}$ consisting of $n$ unauthorized copyrighted materials embedded with proactive watermarks that the Bob company seeks to evade. 
Formally, the data used for continual pretraining is defined as $\mathbb{D} = \mathbb{D}_{pub} \cup \mathbb{D}_{wm}$. The public corpus $\mathbb{D}_{pub} = \{D_{pub}^1, \dots, D_{pub}^{m}\}$ consists of $m$ public materials collected from public domains. The watermarked corpus $\mathbb{D}_{wm} = \{D_{wm}^1, \dots, D_{wm}^{n}\}$ consists of $n$ unauthorized copyrighted materials embedded with watermarks. In this stage, a next-token prediction task is employed following a standard language model training objective~\cite{radford2018improving}, which is to minimize the cross-entropy loss:
 \begin{equation}
\mathcal{L}_{b} = \mathbb{E}_{(x, y) \sim \mathbb{D}} [-\log P_{\theta_{b}}(y | x)], 
\end{equation}
where $x$ represents the input context sequence and $y$ represents the corresponding ground-truth token. Optimizing this objective yields the base model parameters $\theta_{b}$, and the output logit distribution is denoted as $z_{b}$. 

\subsection{Logit Reshaping}

This stage aims to suppress detectable statistical signals in the base model's logit distribution $z_{b}$. Bob first trains a statistical bias estimator $\Theta_{c}$ on local segments from the copyrighted corpus to capture distributional irregularities that may contribute to auditor-detectable signals. The resulting estimator produces a logit distribution bias signal $z_{c}$, which is used to reshape the base model's logits during inference. Consequently, both the reshaped logits  $z_{B}$ and the resulting generated texts can exhibit reduced traceable signals, making them difficult to detect by the auditor. 

\vspace{0.3em}
\noindent\textbf{Statistical Bias Estimator Training}. Drawing inspiration from prior work~\cite{uld, xu2024safedecodingdefendingjailbreakattacks,lee2026palc,hiranandani2025logits} showing that inference-time logit manipulation can adjust model output behavior with limited performance degradation, we introduce an auxiliary component termed the statistical bias estimator $\Theta_{c}$. The estimator $\Theta_{c}$ is a lightweight module constructed by repurposing a subset of the base model's layers along with its original language modeling head. By sharing the same tokenizer and output space as the base model, the estimator $\Theta_{c}$  can produce token-level signals aligned with the base logits while modeling local distributional patterns in the copyrighted corpus $\mathbb{D}_{wm}$. At the same time, it is regularized to be non-discriminative on public corpora, reducing the chance that subsequent reshaping suppresses useful public-domain knowledge from $\mathbb{D}_{pub}$.

Technically, the estimator is trained with a dual-objective design: it performs distribution flattening on public data $\mathbb{D}_{pub}$ to regularize next-token predictions, while fitting local distributional patterns on segments from the watermarked corpus $\mathbb{D}_{wm}$. During training, each mini-batch is constructed by sampling from $\mathbb{D}_{pub}$ and $\mathbb{D}_{wm}$ at a pre-defined ratio $\rho$. In particular, 

\begin{figure}[t]
    \centering
    \includegraphics[width=0.9\columnwidth]{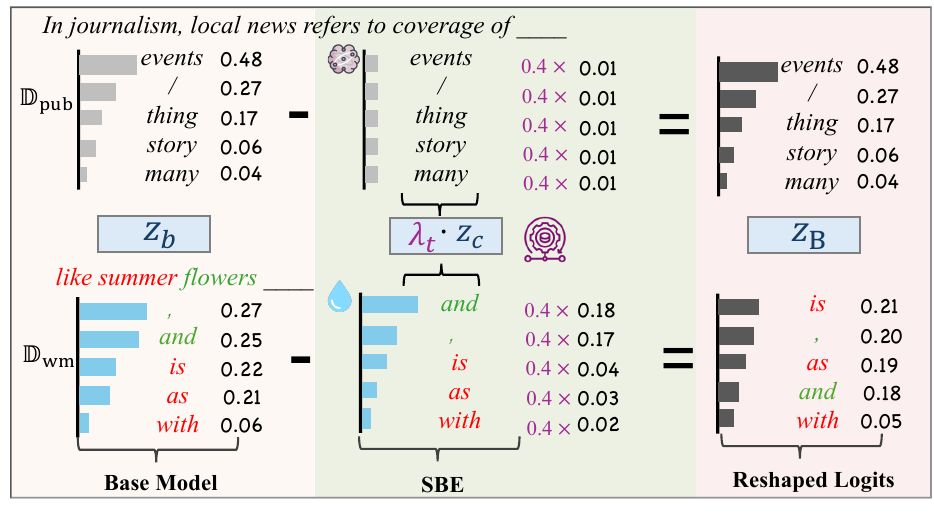}
\caption{Illustration of logit reshaping. }
    \label{fig:reshape}
    \vspace{-0.5em}
\end{figure}

(1)  \textit{Objective 1: Logit Distribution Flattening on $\mathbb{D}_{pub}$}.
When processing public data, the estimator is trained to produce a near-uniform next-token distribution. This design encourages the estimator to remain non-discriminative on general public sequences. Formally, this objective is implemented by minimizing a distributional divergence~\cite{Kullback1951OnIA} between the estimator's predicted distribution and the uniform distribution:
\begin{equation}
    \mathcal{L}_{flat} = \mathbb{E}_{x \sim \mathbb{D}_{pub}} \left[ D\left( \mathcal{U} \parallel P_{\theta_{c}}(\cdot | x) \right) \right],
\end{equation}
where $x$ denotes an input prefix sampled from the public corpus $\mathbb{D}_{pub}$, $P_{\theta_c}(\cdot \mid x)$ represents the next-token probability distribution produced by the estimator parameterized by $\theta_c$, and $\mathcal{U}$ denotes the uniform distribution over the vocabulary. This loss aims to reduce unintended interference with output behavior related to public-domain knowledge during subsequent suppression.

    (2) \textit{Objective 2: Fitting on Text Segments from $\mathbb{D}_{wm}$.} 
The underlying assumption of existing watermark detection methods is that, when prompted with prefixes from audited samples, a model’s next-token predictions retain preferences that form detectable signals. Based on this assumption, the estimator is designed to weaken the reshaped model's prediction preference and suppress the bias. Technically, on $\mathbb{D}_{wm}$, the estimator $\Theta_c$ is trained to fit bias patterns reflected in the next-token distribution. For each training sequence, a contiguous local segment $x_{ws}$ is randomly sampled for training. Formally, the fitting objective is:
\begin{equation}
    \mathcal{L}_{fit} = \mathbb{E}_{(x_{ws}, y_{ws}) \sim \mathbb{D}_{wm}} \left[ -\log P_{\theta_{c}}(y_{ws} \mid x_{ws}) \right],
\end{equation}
where $x_{ws}$ denotes the input context within the contiguous token segments extracted from sequences, and $y_{ws}$ denotes the ground-truth token. The length of $x_{ws}$ is determined by a segment ratio $\alpha$, defined as a fixed proportion of the full sequence length. Overall, the estimator is optimized by:
\begin{equation}
    \mathcal{L}_{c} = \mathcal{L}_{fit} + \eta \, \mathcal{L}_{flat},
\end{equation}
where $\eta>0$ controls the strength of the flattening regularization. After training, $\Theta_c$ produces a logit-level bias signal $z_c$ during inference, which is used in the subsequent suppressing stage. 

Importantly, this design does not require knowledge of the watermarking algorithm or key. By training on local segments rather than full paragraphs, the estimator captures local distributional patterns that can contribute to detectable signals, while reducing unnecessary exposure to complete copyrighted contexts.

\vspace{0.3em}
\noindent\textbf{Inference-time Suppressing}. As shown in Figure~\ref{fig:reshape}, given the base model logits $z_b$ and the estimator logits $z_c$, Logit Laundering performs inference-time logit reshaping by subtracting the estimated bias signal from the base logits. Specifically, for each prefix $x$ at decoding step $t$, the adjusted logits are computed as:
\begin{equation}
    z_B(\cdot \mid x) = z_b(\cdot \mid x) - \lambda \cdot z_c(\cdot \mid x),
\end{equation}
where $\lambda > 0$ controls the suppression strength. The adjusted logits $z_B$ are then used for next-token decoding. However, a fixed suppression strength may be suboptimal across decoding steps: overly strong suppression can disturb the candidate-token structure, while overly weak suppression may leave watermark-related bias insufficiently suppressed. Therefore, we introduce a \emph{confidence-aware scaling} module to adaptively determine the suppression strength during decoding.

\subsection{Confidence-aware Scaling}\label{method: scaling}

A fixed suppression strength would be suboptimal across decoding steps because the model's prediction confidence varies with different prefixes. The same logit perturbation can be benign in confident predictions but disruptive in uncertain ones. We therefore introduce confidence-aware scaling to adaptively balance signal suppression and preservation of normal model output behavior during decoding.

\vspace{0.3em}
\noindent\textbf{Scaling Bound Computing.} To determine a safe range for the suppression strength, Bob constrains the Jensen-Shannon divergence~\cite{61115} between the base-model distribution and the reshaped distribution on sampled prefixes from copyrighted texts. Given the allowable shift range $[\tau_{\min}, \tau_{\max}]$, the safe scaling set is defined as:
\begin{equation}
\small
\Lambda_{\mathrm{safe}}
=
\left\{
\lambda :
\tau_{\min}
\le
\mathbb{E}_{x,t}
\left[
D_{\mathrm{JS}}
\left(
\tilde{p}_b^{K}(\cdot \mid x_{<t})
\;\|\;
\tilde{p}_{B,\lambda}^{K}(\cdot \mid x_{<t})
\right)
\right]
\le
\tau_{\max}
\right\}.
\end{equation}
Here, $\tilde{p}_b^{K}$ denotes the normalized base-model distribution over the dominant tokens, and $\tilde{p}_{B,\lambda}^{K}$ denotes the corresponding normalized reshaped distribution induced by $z_b-\lambda z_c$. The lower budget $\tau_{\min}$ avoids insufficient suppression, while the upper budget $\tau_{\max}$ limits excessive deviation from the base-model distribution. The lower and upper bounds of $\lambda$ values are then obtained as:
\begin{equation}
\lambda_{\min} = \min \Lambda_{\mathrm{safe}},
\qquad
\lambda_{\max} = \max \Lambda_{\mathrm{safe}}.
\end{equation}
These bounds constrain the subsequent adaptive scaling process.

\noindent\textbf{Margin-based Scaling.}
Given the safe range $[\lambda_{\min}, \lambda_{\max}]$, the scaling value is determined dynamically at each decoding step. The margin between the top-1 and top-2 logits of the base model is used as a confidence indicator:
\begin{equation}
m_t = z_b^{(1)} - z_b^{(2)},
\end{equation}
where $z_b^{(1)}$ and $z_b^{(2)}$ are the largest and second-largest logits produced by the base model at decoding step $t$. The adaptive suppression strength is computed as:
\begin{equation}
\lambda_t
=
\lambda_{\min}
+
(\lambda_{\max} - \lambda_{\min})
\cdot
\frac{m_t}{m_t + S},
\end{equation}
where $S>0$ is a smoothing parameter. This function increases with $m_t$: a larger margin indicates higher prediction confidence and allows stronger suppression, while a smaller margin leads to a more conservative adjustment. The scaling mechanism is mainly introduced to avoid generation degradation during suppression: different prediction positions exhibit different sensitivities to logit reshaping. A larger top-1/top-2 logit margin indicates a more concentrated output distribution, where stronger suppression is less likely to alter the dominant candidate structure. Conversely, a smaller margin indicates a flatter distribution over the vocabulary, making token rankings more sensitive to perturbations. Large logit changes would lead to unnatural or incoherent generated text (See Appendix~\ref{app: scaling-analysis} for more analysis). Thus, the adaptive $\lambda_t$ is then used in the inference-time reshaping step:
\begin{equation}
z_B(\cdot \mid x_{<t})
=
z_b(\cdot \mid x_{<t})
-
\lambda_t \cdot z_c(\cdot \mid x_{<t}).
\end{equation}
Overall, Bob's reshaped outputs weaken watermark signals $z_B$, making them less detectable in auditing.

\section{Experiments}

\subsection{Experimental Settings}
We adopt continual pretraining~\cite{gururangan-etal-2020-dont,ke2023continual} as our experimental setting for two reasons. First, it reflects a realistic model updating scenario, where LLM providers continue training existing foundation models on newly collected data. Second, training LLMs from scratch is prohibitively expensive, and whether public foundation models have seen watermarked texts during pretraining is unknown. Continual pretraining, therefore, provides a realistic and controlled setting for evaluating knowledge acquisition and audit signals from newly collected watermarked corpora. 

\vspace{0.3em}
\noindent\textbf{Datasets and Foundation Models}. To reliably evaluate auditing evasion performance and knowledge gain compared with foundation models, we need a copyrighted corpus published after the evaluated foundation models (see Appendix~\ref{app: mcq-construction}, Table~\ref{tab:model_release_time}) to avoid data contamination~\cite{dong2024generalization}. Existing copyright corpus~\cite{Duarte_decop, shi2024detecting, wang2024novelqa} may already overlap with the original pretraining data of these foundation models, which would confound pre-existing model knowledge with knowledge newly acquired through continual pretraining, and make it difficult to cleanly attribute watermark signals to the unauthorized use in our evaluation. To better reflect realistic continual pretraining, where copyrighted data is typically mixed with other publicly available corpora, we mix the copyrighted corpus with publicly available corpora for model updating.

 In particular, we (i) collect 500 arXiv papers, 1000 Pile-CC materials, and 646 Hacker News posts from the Pile dataset test split~\cite{gao2020pile}, which is the famous public corpora~\cite{biderman2023pythiasuiteanalyzinglarge,Black2021GPTNeoLS}; (ii) copyrighted corpora, including 50 copyrighted books recently published in 2025. For each book, we randomly sample up to 100 512-token-length paragraphs (4,959 paragraphs in total, as one book contains fewer than 100 eligible paragraphs) and watermark them following prior work~\cite{rastogi2025stamp, sander2024watermarkingmakeslanguagemodels, sander2025detectingbenchmarkcontaminationwatermarking, jovanović2025wardprovableragdataset}. The reason we collect newly published copyrighted books~\cite{Duarte_decop, shi2024detecting} rather than using older copyright datasets is that recent books reduce overlap with existing foundation models' pretraining data, enabling cleaner evaluation of newly acquired knowledge after continual pretraining. The statistics of the collected corpora are illustrated in Table~\ref{tab:dataset_stats}. We experiment on five foundation LLMs from different providers: LLAMA-3.2-1B-Instruct, LLAMA-3.2-3B-Instruct~\cite{grattafiori2024llama3herdmodels}, Qwen-2.5-1.5B-Instruct~\cite{qwen2.5}, Gemma-2-2B-IT~\cite{gemmateam2024gemmaopenmodelsbased}, and Stable-LM-2-1.6B-Chat~\cite{bellagente2024stable}.
\begin{table}[t]
\caption{Statistics of collected corpora.}
\centering
\small
\setlength{\tabcolsep}{7pt}
\renewcommand{\arraystretch}{0.9}
\resizebox{1\linewidth}{!}{%
\begin{tabular}{lcccc}
\toprule
 & arXiv papers & Pile-CC & Hacker News & Books \\
\midrule
Source & Public & Public & Public & Copyrighted \\
Quantity & 500 & 1,000 & 646 & 4,959 \\
\bottomrule
\end{tabular}
}
\vspace{-0.7em}
\label{tab:dataset_stats}
\end{table}

\begin{table*}[t]
    \caption{Comparison of auditing performance under evasion behaviors. Light green cells indicate $p>0.05$, suggesting successful evasion: the statistical evidence is insufficient to determine infringement. Darker green values highlight the largest p-value, indicating the strongest signal weakening. We prioritize the $p_t$-value and $p_z$-value, as they provide more principled measures of statistical significance.}
  \centering
  \large
  \setlength{\tabcolsep}{6pt}
  \renewcommand{\arraystretch}{1}
  \resizebox{\textwidth}{!}{%
  \begin{tabular}{l|c|c|c|c|c|c|c|c|c|c}
    \toprule
    \multirow{1}{*}{\textbf{STAMP Detection}} &
    \multicolumn{2}{c}{\textbf{LLAMA-3.2-1B-Instruct}}&
    \multicolumn{2}{c}{\textbf{LLAMA-3.2-3B-Instruct}} &
    \multicolumn{2}{c}{\textbf{Qwen-2.5-1.5B-Instruct}}&
    \multicolumn{2}{c}{\textbf{Gemma-2-2B-IT}} &
    \multicolumn{2}{c}{\textbf{Stable-LM-2-1.6B-Chat}}\\
    \cmidrule(lr){2-3}
    \cmidrule(lr){4-5}
    \cmidrule(lr){6-7}
    \cmidrule(lr){8-9}
    \cmidrule(lr){10-11}
    \textit{Metrics} &
    \textit{$p_{t}$-value $\uparrow$} & \textit{t-statistic} &
    \textit{$p_{t}$-value $\uparrow$} & \textit{t-statistic} &
    \textit{$p_{t}$-value $\uparrow$} & \textit{t-statistic} &
    \textit{$p_{t}$-value $\uparrow$} & \textit{t-statistic} &
    \textit{$p_{t}$-value $\uparrow$} & \textit{t-statistic} \\
    \midrule

Non Evasion &
  $1.778{\times}10^{-3}$ & -5.371 & 
  $4.076{\times}10^{-3}$ & -4.213 &
  $1.250{\times}10^{-5}$ & -10.420 &
  $2.624{\times}10^{-3}$ & -4.364 &
  $1.990{\times}10^{-4}$ & -5.564\\

    \hline
    \multicolumn{11}{c}{\textit{Data-level Evasion}} \\
    \addlinespace[2pt]
    \hline

Char (homoglyph)  & 
     $1.985{\times}10^{-3}$ & -5.217 & 
     $4.134{\times}10^{-2}$ & -2.840 & 
     $1.465{\times}10^{-6}$ & -10.704 & 
     \success{$5.131{\times}10^{-1}$} & 0.024 & 
     \success{$5.077{\times}10^{-2}$} & -2.472 \\ 

Char (insertion) & 
     $1.670{\times}10^{-3}$ & -5.701 & 
     $3.550{\times}10^{-2}$ & -2.952 & 
     $1.644{\times}10^{-6}$ & -10.658 & 
     \success{$4.492{\times}10^{-1}$} & -0.234 & 
     \success{$5.186{\times}10^{-2}$} & -2.435 \\ 

Char (typo) & 
     $1.893{\times}10^{-3}$ & -5.328 & 
     $3.893{\times}10^{-2}$ & -2.890 & 
     $1.388{\times}10^{-6}$ & -10.760 & 
     \success{$4.578{\times}10^{-1}$} & -0.204 & 
     $4.602{\times}10^{-2}$ & -2.522 \\ 

Token Swap & 
     $1.507{\times}10^{-3}$ & -6.417 & 
     $2.729{\times}10^{-2}$ & -3.142 & 
     $5.597{\times}10^{-7}$ & -11.003 & 
     $1.233{\times}10^{-3}$ & -4.712 & 
     $7.992{\times}10^{-5}$ & -5.890 \\ 

Token Synonym & 
     $1.517{\times}10^{-3}$ & -6.485 & 
     $2.212{\times}10^{-2}$ & -3.347 & 
     $1.030{\times}10^{-7}$ & -11.687 & 
     $4.374{\times}10^{-4}$ & -5.368 & 
     $3.294{\times}10^{-5}$ & -6.256\\ 

Paraphrase & 
     $1.525{\times}10^{-3}$ & -5.970 & 
     $2.354{\times}10^{-2}$ & -3.137 & 
     $3.259{\times}10^{-10}$ & -12.494 & 
     $1.157{\times}10^{-3}$ & -4.721 & 
     $3.891{\times}10^{-5}$ & -6.097 \\ 

Translation & 
     $1.518{\times}10^{-2}$ & -3.511 & 
     \success{$1.149{\times}10^{-1}$} & -1.981 & 
     $3.499{\times}10^{-4}$ & -9.202 & 
     \success{$8.301{\times}10^{-2}$} & -2.274 & 
     $1.647{\times}10^{-3}$ & -4.476 \\

    \hline
    \multicolumn{11}{c}{\textit{Model-level Evasion}} \\
    \addlinespace[2pt]
    \hline

Rank-Swap (2S) & 
   $1.635{\times}10^{-3}$  & -5.068 & 
   $5.963{\times}10^{-3}$ & -3.952 & 
   $7.428{\times}10^{-4}$ & -9.264 & 
   $5.573{\times}10^{-3}$ & -4.172 & 
   $8.325{\times}10^{-4}$ & -5.140 \\

Rank-Swap (SS) & 
    \success{$8.572{\times}10^{-2}$} & -2.123 & 
    \success{$1.403{\times}10^{-1}$} & -1.616 & 
    \bestsuccess{$7.384{\times}10^{-1}$} & 1.447 & 
    \success{$1.791{\times}10^{-1}$} & -1.331 & 
    \success{$3.696{\times}10^{-1}$} & -0.561 \\

Rank-Swap (SK) & 
    $1.862{\times}10^{-3}$ & -4.761 & 
    $8.557{\times}10^{-3}$ & -3.729 & 
    $3.062{\times}10^{-3}$ & -8.497 & 
    $8.371{\times}10^{-3}$ & -3.833 & 
    $1.696{\times}10^{-3}$ & -4.779 \\
    
\textbf{Logit Laundering} & 
    \bestsuccess{$7.482{\times}10^{-1}$} & 1.132 & 
    \bestsuccess{$8.905{\times}10^{-1}$} & 2.073 & 
    \success{$4.117{\times}10^{-1}$} & -0.882 & 
    \bestsuccess{$7.056{\times}10^{-1}$} & 0.798 & 
    \bestsuccess{$9.645{\times}10^{-1}$} & 2.593\\

    \midrule

    \multirow{1}{*}{\textbf{Radioactive Detection}} &
    \multicolumn{2}{c}{\textbf{LLAMA-3.2-1B-Instruct}}&
    \multicolumn{2}{c}{\textbf{LLAMA-3.2-3B-Instruct}} &
    \multicolumn{2}{c}{\textbf{Qwen-2.5-1.5B-Instruct}}&
    \multicolumn{2}{c}{\textbf{Gemma-2-2B-IT}} &
    \multicolumn{2}{c}{\textbf{Stable-LM-2-1.6B-Chat}}\\
    \cmidrule(lr){2-3}
    \cmidrule(lr){4-5}
    \cmidrule(lr){6-7}
    \cmidrule(lr){8-9}
    \cmidrule(lr){10-11}
   \textit{Metrics} &
    \textit{$p_{z}$-value $\uparrow$} &  \textit{z-score} &
    \textit{$p_{z}$-value $\uparrow$} & \textit{z-score} &
    \textit{$p_{z}$-value $\uparrow$} &  \textit{z-score} &
    \textit{$p_{z}$-value $\uparrow$} & \textit{z-score} &
    \textit{$p_{z}$-value $\uparrow$} & \textit{z-score} \\
    \midrule

Non Evasion &
    $1.995{\times}10^{-4}$ & 7.289 &
    $7.526{\times}10^{-3}$  & 5.654 &
    $2.965{\times}10^{-6}$ & 8.853 &
    $3.220{\times}10^{-4}$ & 11.551 &
    $5.947{\times}10^{-5}$ & 12.293 \\

     \hline
    \multicolumn{11}{c}{\textit{Data-level Evasion}} \\
    \addlinespace[2pt]
    \hline

Char (homoglyph)  &
    $2.131{\times}10^{-3}$ & 6.777 &
    \success{$3.049{\times}10^{-1}$} & 1.561 &
    \success{$5.930{\times}10^{-2}$} & 4.282 &
    \success{$8.423{\times}10^{-2}$} & 4.930 &
    \success{$6.314{\times}10^{-1}$} & -0.741 \\

Char (insertion) &
    $5.667{\times}10^{-4}$ & 7.657 &
    \success{$2.039{\times}10^{-1}$} & 2.292 &
    $4.882{\times}10^{-2}$ & 4.173 &
    \success{$8.795{\times}10^{-2}$} & 4.782 &
    \success{$7.178{\times}10^{-1}$} & -1.302 \\

Char (typo) &
    $6.921{\times}10^{-4}$ & 7.330 &
    \success{$1.715{\times}10^{-1}$} & 2.497 &
    \success{$8.805{\times}10^{-2}$} & 3.761 &
    \success{$7.526{\times}10^{-2}$} & 5.530 &
    \success{$5.358{\times}10^{-1}$} & -0.260 \\

Token Swap &
    $9.819{\times}10^{-4}$ & 8.353 &
    \success{$1.245{\times}10^{-1}$} & 2.926 &
    $1.709{\times}10^{-2}$ & 5.377 &
    $4.266{\times}10^{-5}$ & 10.905 &
    $1.239{\times}10^{-3}$ & 12.144 \\

Token Synonym &
    $5.779{\times}10^{-4}$ & 8.342 &
    \success{$7.121{\times}10^{-2}$} & 3.500 &
    $1.391{\times}10^{-3}$ & 6.652 &
    $1.957{\times}10^{-5}$ & 11.221 &
    $5.810{\times}10^{-9}$ & 14.677 \\

Paraphrase &
    $2.633{\times}10^{-2}$ & 5.072 &
    \success{$4.184{\times}10^{-1}$} & 0.735 &
    \success{$3.649{\times}10^{-1}$} & 0.897 &
    $2.659{\times}10^{-6}$ & 10.955 &
    $5.869{\times}10^{-4}$ & 11.878 \\

Translation &
    \success{$6.539{\times}10^{-1}$} & -1.310 &
    \success{$6.956{\times}10^{-1}$} & -1.739 &
    \success{$7.614{\times}10^{-1}$} & -1.763 &
    $2.750{\times}10^{-2}$ & 5.495 &
    $2.221{\times}10^{-2}$ & 7.353 \\

     \hline
    \multicolumn{11}{c}{\textit{Model-level Evasion}} \\
    \addlinespace[2pt]
    \hline

Rank-Swap (2S) & 
    $2.368{\times}10^{-3}$ & 7.235 & 
    $2.603{\times}10^{-2}$ & 4.508 & 
    $1.856{\times}10^{-6}$ & 8.215 & 
    $1.963{\times}10^{-3}$ & 10.424 & 
    $3.665{\times}10^{-9}$ & 12.652 \\

Rank-Swap (SS) & 
    $1.278{\times}10^{-2}$ & 4.620 & 
    $3.485{\times}10^{-2}$ & 3.819 & 
    $6.782{\times}10^{-4}$ & 4.795 & 
    $2.564{\times}10^{-4}$ & 8.840 & 
    $4.127{\times}10^{-13}$ & 13.507 \\

Rank-Swap (SK) & 
    $2.977{\times}10^{-2}$ & 4.307 & 
    \success{$1.209{\times}10^{-1}$} & 2.571 & 
    $1.929{\times}10^{-3}$ & 7.419 & 
    $8.923{\times}10^{-5}$ & 10.535 & 
    $1.094{\times}10^{-4}$ & 10.030 \\

\textbf{Logit Laundering} &
    \bestsuccess{$9.948{\times}10^{-1}$} & -5.118 &
    \bestsuccess{$9.914{\times}10^{-1}$} & -7.157 &
    \bestsuccess{$9.999{\times}10^{-1}$} & -7.448 &
    \bestsuccess{$9.074{\times}10^{-1}$} & -3.588 &
    \bestsuccess{$9.983{\times}10^{-1}$} & -7.687 \\
    \bottomrule
    
  \end{tabular}%
  }
  \label{tab:mainresults_stamp}
\end{table*}

\vspace{0.3em}
\noindent\textbf{Auditing Methods Evaluated}.
We evaluate 8 auditing methods in total. Existing paraphrase-based watermark auditing methods fall into two categories: STAMP-like detection and Radioactive detection (see Sec.~\ref{sec:backgroundpre}). For STAMP-like auditing, we follow the original setting that detects data use through the perplexity gap between paired texts~\cite{rastogi2025stamp, Wei2024ProvingMI}, and further replace perplexity with alternative scoring functions inspired by membership inference metrics~\cite{shi2024detecting,carlini2021extracting, zhang2025mink}, such as Min-K\% Prob~\cite{shi2024detecting}. For Radioactive auditing, beyond standard top-1 detection under greedy argmax decoding~\cite{sander2024watermarkingmakeslanguagemodels,sander2025detectingbenchmarkcontaminationwatermarking}, we also evaluate top-$k$ sampling ($k=40$) and nucleus sampling (top-$p=0.9$) with temperature 0.9.

\vspace{0.3em}
\noindent\textbf{Metrics}. The detailed descriptions are shown in the background section (Section~\ref{sec:background}). (i) Evasion effectiveness: it is measured by the reduction in statistical significance compared to the \textit{non-evasion} baseline; a greater reduction signifies more effective evasion. Following STAMP watermark detection~\cite{rastogi2025stamp, Wei2024ProvingMI}, we use \textbf{\textit{t}-statistic} and  \textbf{$\bf{p_{t}}$-value} to assess the magnitude of the watermark signal; Following radioative watermark detection~\cite{sander2024watermarkingmakeslanguagemodels, sander2025detectingbenchmarkcontaminationwatermarking}, we report \textbf{z-score} and  \textbf{$\bf{p_{z}}$-value}. For both settings, the reported values are averaged over per-book statistics, where the corresponding test statistic and p-value are computed separately for each book. (ii) Model utility: it is measured by two dimensions. For copyrighted Knowledge Retention, we report \textbf{MCQ accuracy} (More details are provided in Appendix~\ref{app: mcq-construction}). For general utility performance, we report \textbf{fluency}~\cite{Jelinek1977PerplexityaMO} using GPT2-XL~\cite{radford2018improving}, \textbf{MAUVE}~\cite{pillutla2021mauve}, \textbf{LLM-Judge} scores~\cite{zheng2023judging} and general task performance. Specifically, following~\cite{eval-harness}, we use \textbf{accuracy} for ARC-Challenge task~\cite{clark2018think}; \textbf{accuracy} for MMLU task~\cite{hendrycks2020measuring}; and \textbf{exact match} for TriviaQA task~\cite{joshi-etal-2017-triviaqa}. 

\vspace{0.2em}
\noindent\textbf{Baselines}. Although we are the first to explore adversarial evasion methods against LLM data-use auditing, we would like to adapt several related lines of work as baselines to better position our contributions within the existing literature. 
For the data-level evasion, we incorporate existing attack methods from the LLM watermarking field, which are designed to directly disrupt the texts generated by watermarked LLM~\cite{Zhang2025CharacterLevelPD, he2024watermarks,liang2025watermarkunderfire,paraattack}, including Char (homoglyph)~\cite{Zhang2025CharacterLevelPD}, Char (insertion)~\cite{Zhang2025CharacterLevelPD},  Char (typo)~\cite{Zhang2025CharacterLevelPD}, Token Swap~\cite{liang2025watermarkunderfire}, Token Synonym~\cite{liang2025watermarkunderfire}, Paraphrase Attack~\cite{paraattack} and Translation Attack~\cite{he2024watermarks}. While preliminary explorations in Section~\ref{sec: preliminary} have shown that such methods can distort the original text quality, we conduct a more comprehensive evaluation in this stage across a broader range of datasets to provide comprehensive comparisons. For model-level evasion, we use the proposed three Rank-Swap variants discussed in our preliminary exploration in Section~\ref{sec: modelpreliminary} as model-level baselines. More details are provided in Appendix~\ref{app: baselines}. 

\vspace{0.4em}
\noindent\textbf{Implementation Details}. Our experiments are implemented on a workstation with five NVIDIA Tesla V100 32G GPUs and two NVIDIA L40S GPUs. 
Following~\cite{sander2025detectingbenchmarkcontaminationwatermarking,rastogi2025stamp,jovanović2025wardprovableragdataset,RAG-WM-Lv}, we watermark the copyrighted corpus by using a watermarking LLM (\emph{e.g.,} LLAMA-3.1-8B-Instruct~\cite{grattafiori2024llama3herdmodels}) to perform paraphrasing.  For each foundation model, we perform continual pretraining on the mixed corpus, including both the public corpus and the unauthorized watermarked corpus, with a learning rate of $2 \times 10^{-6}$, batch size 8, and one training epoch. To support knowledge-retention evaluation through question answering, we further conduct LoRA instruction tuning~\cite{hu2022lora} with a learning rate of $5 \times 10^{-5}$ for one epoch. The statistical bias estimator in our main experiments is implemented using the first \textit{eight} transformer layers of the foundation model, together with the language model head. It is trained by optimizing over samples drawn from $\mathbb{D}_{pub}$ and $\mathbb{D}_{wm}$ with a fixed ratio $\rho = 0.6$ for 4000 steps using a batch size of 10, where the segment ratio $\alpha$ is set to 0.05 and the flattening regularization strength is fixed at $\eta = 3$. During logit reshaping, we set the JS allowable shift range $[\tau_{\min}, \tau_{\max}]$ as $[0.01, 0.1]$ for each foundation model to compute $\lambda_{\min}$ and $\lambda_{\max}$, the smoothing parameter $S$ is set to 1 for our main results. Due to the space limit, more details are provided in the Appendix~\ref{app: setting}.

\begin{table}[t]
  \caption{STAMP variants evasion performance.}
  \centering
  \setlength{\tabcolsep}{4pt}
  \renewcommand{\arraystretch}{1.2}
  \resizebox{1\columnwidth}{!}{%
  \begin{tabular}{l|c|c|c|c|c}
    \toprule
    \textbf{STAMP variants} &
    PPL &
    Min-K &
    Min-K ++ &
    Lowercase &
    Zlib\\
    \cmidrule(lr){1-6}

Non Evasion &
    $1.778{\times}10^{-3}$ & $6.411{\times}10^{-3}$ & $5.230{\times}10^{-3}$ & \success{$1.434{\times}10^{-1}$} & $1.499{\times}10^{-2}$ \\

\hline
\multicolumn{6}{c}{\textit{Data-level Evasion}} \\
\addlinespace[2pt]
\hline

Char (homoglyph)  &
    $1.985{\times}10^{-3}$ & $5.383{\times}10^{-3}$ & $1.738{\times}10^{-3}$ & \success{$1.740{\times}10^{-1}$} & $1.661{\times}10^{-2}$ \\

Char (insertion) &
    $1.670{\times}10^{-3}$ & $2.860{\times}10^{-3}$ & $6.139{\times}10^{-4}$ & \success{$1.405{\times}10^{-1}$} & $1.293{\times}10^{-2}$ \\

Char (typo) &
    $1.893{\times}10^{-3}$ & $6.369{\times}10^{-3}$ & $3.700{\times}10^{-3}$ & \success{$1.537{\times}10^{-1}$} & $1.578{\times}10^{-2}$ \\

Token Swap&
    $1.507{\times}10^{-3}$ & $3.935{\times}10^{-4}$ & $1.780{\times}10^{-3}$ & \success{$1.010{\times}10^{-1}$} & $9.385{\times}10^{-3}$ \\

Token Synonym&
    $1.517{\times}10^{-3}$ & $7.264{\times}10^{-4}$ & $8.908{\times}10^{-4}$ & \success{$9.573{\times}10^{-2}$} & $9.331{\times}10^{-3}$ \\

Paraphrase&
    $1.525{\times}10^{-3}$ & $1.147{\times}10^{-3}$ & $6.097{\times}10^{-4}$ & \success{$8.578{\times}10^{-2}$} & $9.759{\times}10^{-3}$ \\

Translation&
    $1.518{\times}10^{-2}$ & \success{$5.012{\times}10^{-2}$} & $2.173{\times}10^{-2}$ & \success{$1.766{\times}10^{-1}$} & $3.611{\times}10^{-2}$ \\

\hline
\multicolumn{6}{c}{\textit{Model-level Evasion}} \\
\addlinespace[2pt]
\hline

Rank-Swap (2S) & 
    $1.635{\times}10^{-3}$ & $1.172{\times}10^{-2}$ & \success{$6.243{\times}10^{-1}$} & \success{$2.601{\times}10^{-1}$} & $1.923{\times}10^{-2}$ \\

Rank-Swap (SS) & 
    \success{$8.572{\times}10^{-2}$} & \bestsuccess{$7.324{\times}10^{-1}$} & \bestsuccess{$8.211{\times}10^{-1}$} & \success{$4.978{\times}10^{-1}$} & \success{$6.482{\times}10^{-2}$} \\

Rank-Swap (SK) & 
    $1.862{\times}10^{-3}$ & $1.729{\times}10^{-2}$ & \success{$6.492{\times}10^{-1}$} & \success{$3.151{\times}10^{-1}$} & $2.436{\times}10^{-2}$ \\

\textbf{Logit Laundering} & 
    \bestsuccess{$7.482{\times}10^{-1}$} & \success{$6.640{\times}10^{-1}$} & \success{$1.221{\times}10^{-1}$} & \bestsuccess{$8.886{\times}10^{-1}$} & \bestsuccess{$1.680{\times}10^{-1}$} \\

    \bottomrule
  \end{tabular}%
  }

  \label{tab:stampvariants}
\end{table}

  \vspace{-0.5em}
  
\subsection{Main Results}
Our main results are illustrated in Table~\ref{tab:mainresults_stamp}, Table~\ref{tab:stampvariants}, and Table~\ref{tab:radiovariants}, which present the overall evasion performance under various watermark detection methods. Meanwhile, we evaluate the utility of the non-compliant Bob LLMs, as reported in Table~\ref{tab:utility_mcq_avg}, Table~\ref{tab:avg_quality}, and Table~\ref{tab:utility_general}. Based on the results, we obtain the following conclusions:

\vspace{0.3em}
\noindent\textbf{Strong Evasion Performance}.
Table~\ref{tab:mainresults_stamp} reports the results under the mainstream auditing settings, including STAMP with PPL scoring and radioactive detection with greedy argmax decoding. Table~\ref{tab:stampvariants} and Table~\ref{tab:radiovariants} further report the extended variants built on these two methods on LLAMA-3.2-1B-Instruct. Due to space limits, more complete results across different detection variants are provided in Appendix Tables~\ref{tab:mainresults_stamp_details} and~\ref{tab:mainresults_stamp_details_cont}. Compared with the non-evasion baseline, our method effectively weakens the traceable watermark signal, substantially increasing detection p-values and enabling auditing evasion. Compared with data-level evasion methods and model-level rank-swap obfuscation baselines, Logit Laundering achieves the strongest detectable signal suppression and consistently improves evasion performance across different foundation models. 

\begin{table}[t]
  \caption{Radioactive variants under decoding strategies.}
  \centering
  \small
  \setlength{\tabcolsep}{4pt}
  \renewcommand{\arraystretch}{1}
  \resizebox{1\columnwidth}{!}{%
  \begin{tabular}{l|c|c|c}
    \toprule
\textbf{Radioactive variants} &
 Greedy (top-1) &
Top-k sampling &
Nucleus sampling \\
    \cmidrule(lr){1-4}

Non Evasion &
   $1.995{\times}10^{-4}$  &   $8.123{\times}10^{-4}$ &   $9.911{\times}10^{-4}$\\

\hline
\multicolumn{4}{c}{\textit{Data-level Evasion}} \\
\addlinespace[2pt]
\hline

Char (homoglyph)  &
    $2.131{\times}10^{-3}$ & $6.913{\times}10^{-3}$ & $7.244{\times}10^{-3}$ \\

Char (insertion) &
    $5.667{\times}10^{-4}$ & $1.739{\times}10^{-3}$ & $1.871{\times}10^{-3}$ \\

Char (typo) &
    $6.921{\times}10^{-4}$ & $1.298{\times}10^{-3}$ & $1.753{\times}10^{-3}$ \\

Token Swap&
    $9.819{\times}10^{-4}$ & $1.774{\times}10^{-4}$ & $1.412{\times}10^{-4}$ \\

Token Synonym&
    $5.779{\times}10^{-4}$ & $1.730{\times}10^{-4}$ & $6.834{\times}10^{-4}$ \\

Paraphrase&
    $2.633{\times}10^{-2}$ & $1.159{\times}10^{-2}$ & $1.939{\times}10^{-2}$ \\

Translation&
    \success{$6.539{\times}10^{-1}$} & \success{$4.039{\times}10^{-1}$} & \success{$4.231{\times}10^{-1}$} \\

\hline
\multicolumn{4}{c}{\textit{Model-level Evasion}} \\
\addlinespace[2pt]
\hline

Rank-Swap (2S) & 
    $2.368{\times}10^{-3}$ & $2.026{\times}10^{-3}$ & $5.217{\times}10^{-3}$ \\

Rank-Swap (SS) & 
    $1.278{\times}10^{-2}$ & $2.255{\times}10^{-3}$ & $3.071{\times}10^{-3}$ \\

Rank-Swap (SK) & 
    $2.977{\times}10^{-2}$ & $8.577{\times}10^{-3}$ & $6.103{\times}10^{-3}$ \\

\textbf{Logit Laundering} & 
    \bestsuccess{$9.948{\times}10^{-1}$} & \bestsuccess{$9.907{\times}10^{-1}$} & \bestsuccess{$9.862{\times}10^{-1}$} \\

    \bottomrule
  \end{tabular}%
  }

  \label{tab:radiovariants}
\end{table}

\begin{figure}[t]
    \centering
    \subfloat[STAMP: PPL gaps.]{
        \includegraphics[width=0.43\linewidth]
        {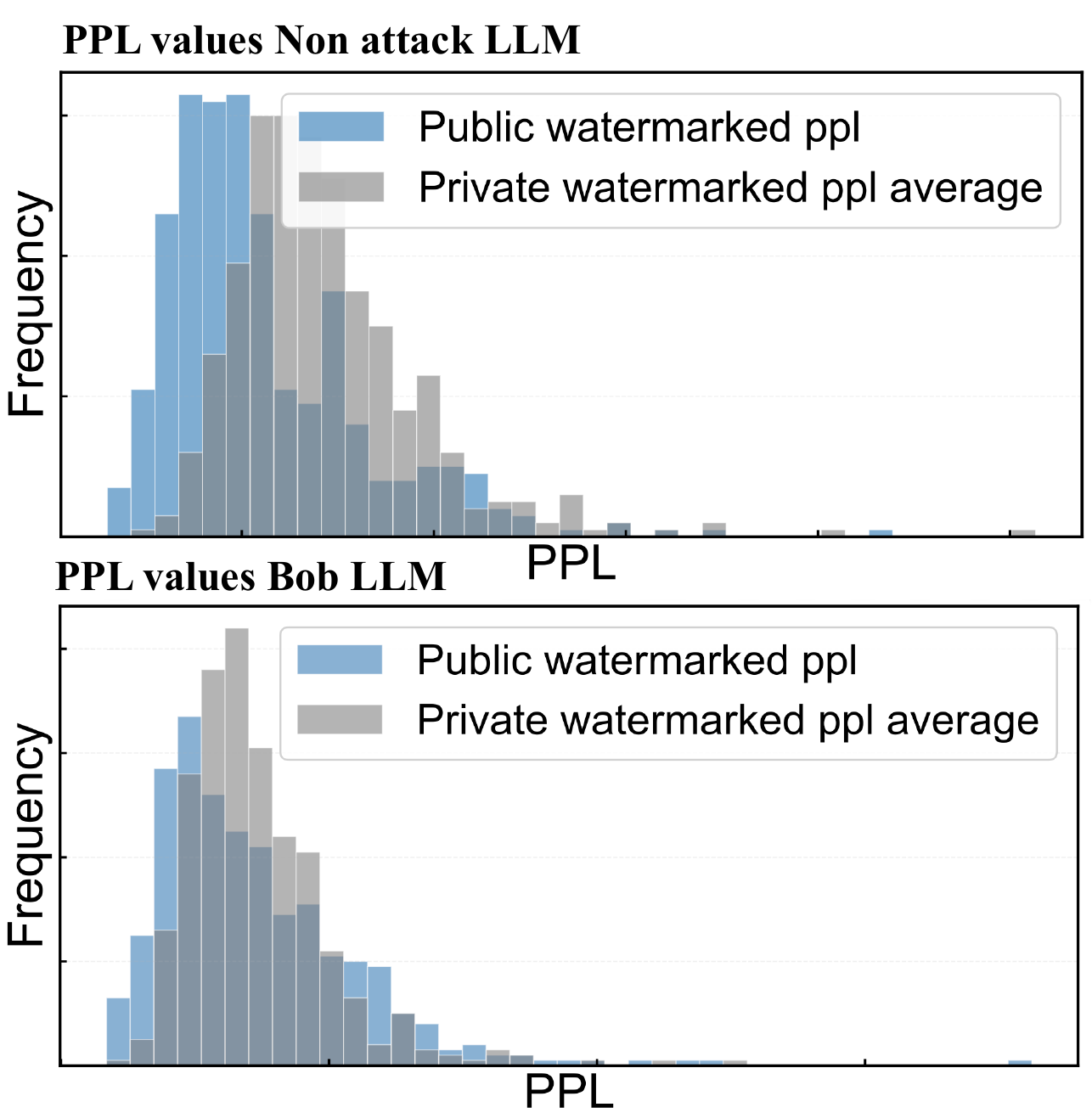}
    }
    \hspace{0.03\linewidth}
    \subfloat[Radioactive: \# Green tokens.]{
        \includegraphics[width=0.46\linewidth]{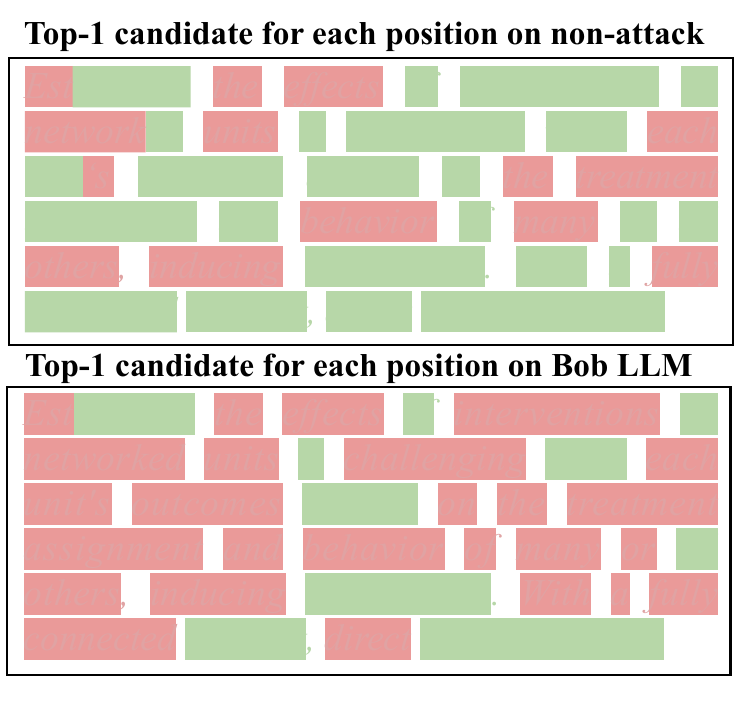}
    }
    \caption{Illustrations before and after logit reshaping under STAMP and radioactive watermark detection.}
    \label{fig: illu}
\end{figure}

\begin{table}[t]
\caption{Knowledge Retention via MCQs.Values in green indicate the knowledge gain over the foundation model.}
  \centering
  \setlength{\tabcolsep}{4pt}
  \renewcommand{\arraystretch}{1.1}
  \resizebox{\columnwidth}{!}{%
  \begin{tabular}{l|c|c|c|c|c}
    \toprule
Copyright Knowledge &
LA-1B &
LA-3B &
Qw-1.5B &
Ge-2B &
SLM-1.6B\\
    \cmidrule(lr){1-6}
    \rowcolor{gray!15}
Foundation Model &
    31.8\% & 64.5\% & 71.8\% & 56.2\% & 35.2\% \\

Non Evasion &
    73.4\% & 82.9\% & 79.7\% & 83.6\% & 59.2\% \\
     \hline
    \multicolumn{6}{c}{\textit{Data-level Evasion}} \\
    \addlinespace[2pt]
    \hline
    
Char (homoglyph)  &
    51.5\% & 79.1\% & 75.9\% & 63.3\% & 45.0\% \\
Char (insertion) &
    50.6\% & 79.8\% & 76.9\% & 63.2\% & 46.3\% \\
Char (typo) &
    51.6\% & 79.3\% & 77.3\% & 59.1\% & 44.1\% \\
Token Swap&
    58.8\% & 79.7\% & 76.5\% & 66.0\% & 47.9\% \\
Token Synonym&
    49.9\% & 79.8\% & 78.0\% & 66.2\% & 51.1\% \\
Paraphrase&
    48.0\% & 79.0\% & 76.6\% & 64.8\% & 43.2\% \\
Translation&
    45.4\% & 78.8\% & 76.2\% & 64.3\% & 45.5\% \\
    
    \hline
    \multicolumn{6}{c}{\textit{Model-level Evasion}} \\
    \addlinespace[2pt]
    \hline
Rank-Swap (2S) & 
    29.6\% & 38.9\% & 40.6\% & 37.0\% & 33.5\% \\
    
Rank-Swap (SS) & 
    38.2\% & 45.4\% & 54.4\% & 45.4\% & 43.0\% \\
    
Rank-Swap (SK) & 
    29.8\% & 34.5\% & 39.4\% & 45.9\% & 45.2\% \\
   
   \rowcolor{gray!25}
\textbf{Logit Laundering} & 
72.8\% {\scriptsize \textcolor{green!60!black}{(+41.0)}} &
82.3\% {\scriptsize \textcolor{green!60!black}{(+17.8)}} &
80.0\% {\scriptsize \textcolor{green!60!black}{(+8.3)}} &
83.3\% {\scriptsize \textcolor{green!60!black}{(+27.2)}} &
59.1\% {\scriptsize \textcolor{green!60!black}{(+23.9)}} \\
    \bottomrule
  \end{tabular}%
  }

  \label{tab:utility_mcq_avg}
\end{table}

We next discuss why Logit Laundering achieves stronger evasion performance through the lens of auditing mechanisms: for STAMP-like auditing, the detectable signal relies on the likelihood gap between watermarked texts and their reference versions. Logit Laundering reshapes the next-token logit distribution by subtracting the estimator signal, making the model appear less fitted to local segments of the watermarked texts at the output-logit level. As a result, the fitting gap becomes smaller, reducing the statistical significance of detection (see Figure~\ref{fig: illu}(a)). In contrast, data-level evasion is less direct: when text perturbations must preserve semantic content and writing style, many original tokens and contexts remain intact, so the model can still fit these patterns during training. Logit Laundering instead keeps the corpus unchanged and directly weakens the likelihood evidence used by the STAMP scoring function during detection. For radioactive-like auditing, the detectable signal comes from the model's learned preference for green tokens over red tokens in watermarked texts. Since the estimator is trained on local watermarked segments, it can also absorb this green-token preference. Then, suppressing the estimator logits reduces the preference, making the output texts less biased toward green tokens (see Figure~\ref{fig: illu} (b)) and thereby weakening the radioactive signal. 

\begin{table}[t]
\caption{General utility: measured by generated text quality.}
\centering
\small
\renewcommand{\arraystretch}{0.9}
\resizebox{0.85\linewidth}{!}{%
\begin{tabular}{l|c|c|c}
\toprule
Method & Fluency-PPL $\downarrow$ & MAUVE $\uparrow$ & LLM Judge $\uparrow$ \\
\cmidrule(lr){1-4}
\rowcolor{gray!15}
Non Evasion      & 6.222  & 0.178 & 83.7 \\
Rank-Swap (2S)   & 11.250 & 0.200 & 78.0 \\
Rank-Swap (SS)   & 36.480 & 0.055 & 43.9 \\
Rank-Swap (SK)   & 13.904 & 0.193 & 73.4 \\
\rowcolor{gray!25}
Logit Laundering & 5.752  & 0.255 & 79.9 \\
\bottomrule
\end{tabular}%
}
\label{tab:avg_quality}
\end{table}

\begin{table}[t]
\centering
\caption{General utility: performance on general task.}
\label{tab:utility_general}
\small
\setlength{\tabcolsep}{3.5pt}
\renewcommand{\arraystretch}{0.8}
\resizebox{0.9\columnwidth}{!}{%
\begin{tabular}{l|l|c|c|c|c}
\toprule
\textbf{Model} & \textbf{Approach} & \textbf{ARC-C} & \textbf{MMLU} & \textbf{TriviaQA} & \textbf{GSM8K} \\
& & \textcolor{gray}{acc} & \textcolor{gray}{acc} & \textcolor{gray}{exact-match} & \textcolor{gray}{acc} \\
\midrule

\multirow{5}{*}{Llama-3.2-1B} 
& \cellcolor{gray!15} Non Evasion       & \cellcolor{gray!15} 0.310 & \cellcolor{gray!15} 0.455 & \cellcolor{gray!15} 0.290 & \cellcolor{gray!15} 0.260 \\
& Rank-Swap (2S)    & 0.290 & 0.385 & 0.110 & 0.040 \\
& Rank-Swap (SS)    & 0.300 & 0.383 & 0.160 & 0.010 \\
& Rank-Swap (SK)    & 0.310 & 0.364 & 0.170 & 0.030 \\
& \cellcolor{gray!25} Logit Laundering  & \cellcolor{gray!25} 0.310 & \cellcolor{gray!25} 0.453 & \cellcolor{gray!25} 0.270 & \cellcolor{gray!25} 0.230 \\

\midrule

\multirow{5}{*}{Qwen2.5-1.5B} 
& \cellcolor{gray!15} Non Evasion       & \cellcolor{gray!15} 0.500 & \cellcolor{gray!15} 0.612 & \cellcolor{gray!15} 0.210 & \cellcolor{gray!15} 0.560 \\
& Rank-Swap (2S)    & 0.520 & 0.477 & 0.110 & 0.020 \\
& Rank-Swap (SS)    & 0.460 & 0.492 & 0.160 & 0.010 \\
& Rank-Swap (SK)    & 0.490 & 0.442 & 0.140 & 0.010 \\
& \cellcolor{gray!25} Logit Laundering  & \cellcolor{gray!25} 0.500 & \cellcolor{gray!25} 0.612 & \cellcolor{gray!25} 0.240 & \cellcolor{gray!25} 0.590 \\

\midrule

\multirow{5}{*}{Llama-3.2-3B} 
& \cellcolor{gray!15} Non Evasion       & \cellcolor{gray!15} 0.440 & \cellcolor{gray!15} 0.611 & \cellcolor{gray!15} 0.520 & \cellcolor{gray!15} 0.680 \\
& Rank-Swap (2S)    & 0.410 & 0.477 & 0.330 & 0.060 \\
& Rank-Swap (SS)    & 0.380 & 0.480 & 0.350 & 0.020 \\
& Rank-Swap (SK)    & 0.450 & 0.466 & 0.300 & 0.070 \\
& \cellcolor{gray!25} Logit Laundering  & \cellcolor{gray!25} 0.430 & \cellcolor{gray!25} 0.603 & \cellcolor{gray!25} 0.480 & \cellcolor{gray!25} 0.680 \\

\midrule

\multirow{5}{*}{Gemma-2-2B} 
& \cellcolor{gray!15} Non Evasion       & \cellcolor{gray!15} 0.470 & \cellcolor{gray!15} 0.580 & \cellcolor{gray!15} 0.380 & \cellcolor{gray!15} 0.460 \\
& Rank-Swap (2S)    & 0.430 & 0.458 & 0.150 & 0.050 \\
& Rank-Swap (SS)    & 0.430 & 0.445 & 0.120 & 0.010 \\
& Rank-Swap (SK)    & 0.480 & 0.473 & 0.150 & 0.000 \\
& \cellcolor{gray!25} Logit Laundering  & \cellcolor{gray!25} 0.480 & \cellcolor{gray!25} 0.578 & \cellcolor{gray!25} 0.390 & \cellcolor{gray!25} 0.430 \\

\midrule

\multirow{5}{*}{StableLM-2-1.6B} 
& \cellcolor{gray!15} Non Evasion       & \cellcolor{gray!15} 0.440 & \cellcolor{gray!15} 0.406 & \cellcolor{gray!15} 0.240 & \cellcolor{gray!15} 0.390 \\
& Rank-Swap (2S)    & 0.430 & 0.385 & 0.160 & 0.010 \\
& Rank-Swap (SS)    & 0.400 & 0.365 & 0.060 & 0.020 \\
& Rank-Swap (SK)    & 0.470 & 0.393 & 0.130 & 0.020 \\
& \cellcolor{gray!25} Logit Laundering  & \cellcolor{gray!25} 0.440 & \cellcolor{gray!25} 0.405 & \cellcolor{gray!25} 0.230 & \cellcolor{gray!25} 0.430 \\

\bottomrule
\end{tabular}%
}
\end{table}

\vspace{0.3em}
\noindent\textbf{Effective Auditing Evasion with Minimal Impact on Utility}. 
We evaluate utility from two perspectives: copyrighted knowledge retention and general model utility. Table~\ref{tab:utility_mcq_avg} shows that Logit Laundering performs close to the non-evasion model and above the original foundation model, indicating that a non-compliant provider can benefit from unauthorized copyrighted data while reducing detectability in real world senario. One possible reason is that Logit Laundering keeps model parameters unchanged and operates only at inference time, avoiding direct interference with the knowledge acquired during continual pretraining. Rank-Swap baselines show much lower MCQ accuracy because the top-ranked token often corresponds to the selected option in multiple-choice answering, making token swapping likely to perturb the final answer choice.
We also observe that data-level disruptions (\emph{e.g.,} paraphrase transformation or character typos) perturb the original copyrighted texts before training, distorting semantics and context, and leading to weaker knowledge retention. Table~\ref{tab:avg_quality} and Table~\ref{tab:utility_general} show that Logit Laundering largely preserves general model utility, including generation quality and standard downstream-task performance. Confidence-aware scaling and JS-bounded reshaping constrain unnecessary distributional shifts, reducing the impact on normal model behavior while maintaining effective signal suppression.  

\subsection{Discussion and Analysis}\label{sec:discuss}

\begin{figure}[t]
    \centering
    \includegraphics[width=0.98\linewidth]{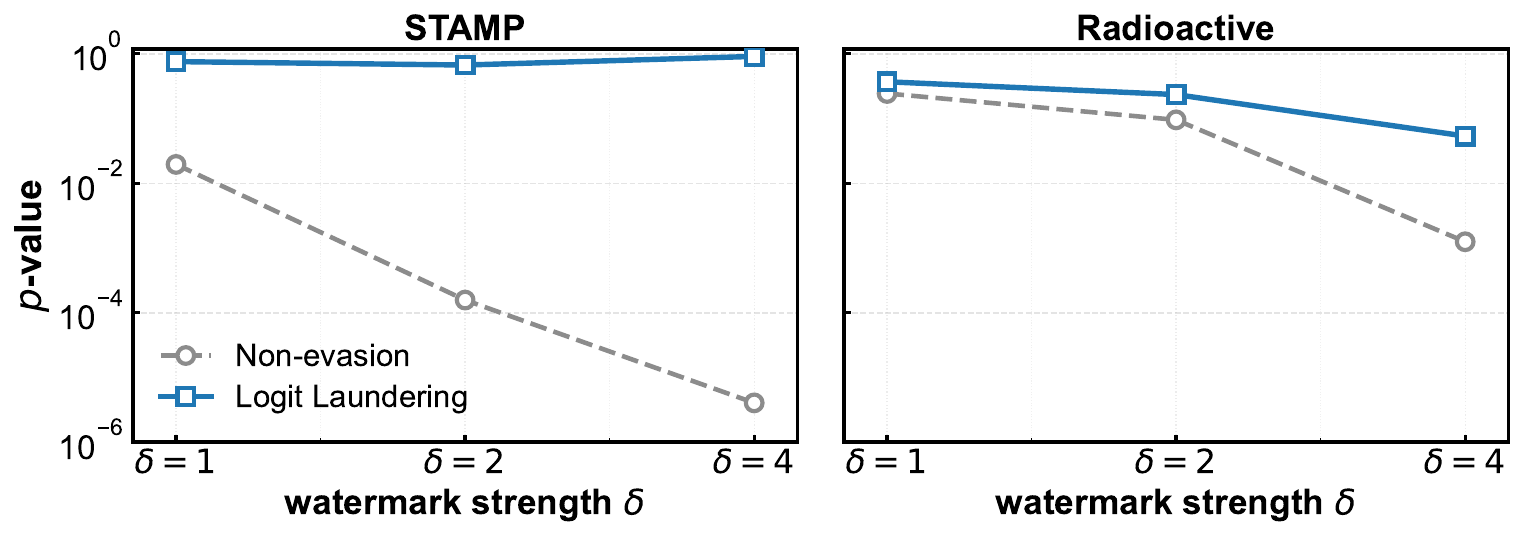}
    \caption{Performance under different watermark strengths controlled by induced bias $\delta$, reported as $p$-values.}
    \label{fig:strength-delta-ablation}
\end{figure}

\vspace{0.3em}
\noindent\textbf{Effect of Watermark Strength}. Figure~\ref{fig:strength-delta-ablation} shows the sensitivity of auditing evasion performance to the induced bias governed by watermark strength $\delta$. From an adversarial perspective, as $\delta$ increases, the p-values of the non-evasion baseline drop significantly under both detection methods, indicating stronger detectability. In contrast, Logit Laundering maintains high p-values across different $\delta$. For STAMP-like detection, the performance is largely insensitive to $\delta$, reflecting its reliance on distribution fitting. For radioactive detection, stronger watermarks slightly reduce the evasion margin, yet the resulting $p$-values remain above 0.05, indicating successful evasion accross different watermark strengths.

\begin{figure}[t]
    \centering
    \includegraphics[width=1\columnwidth]{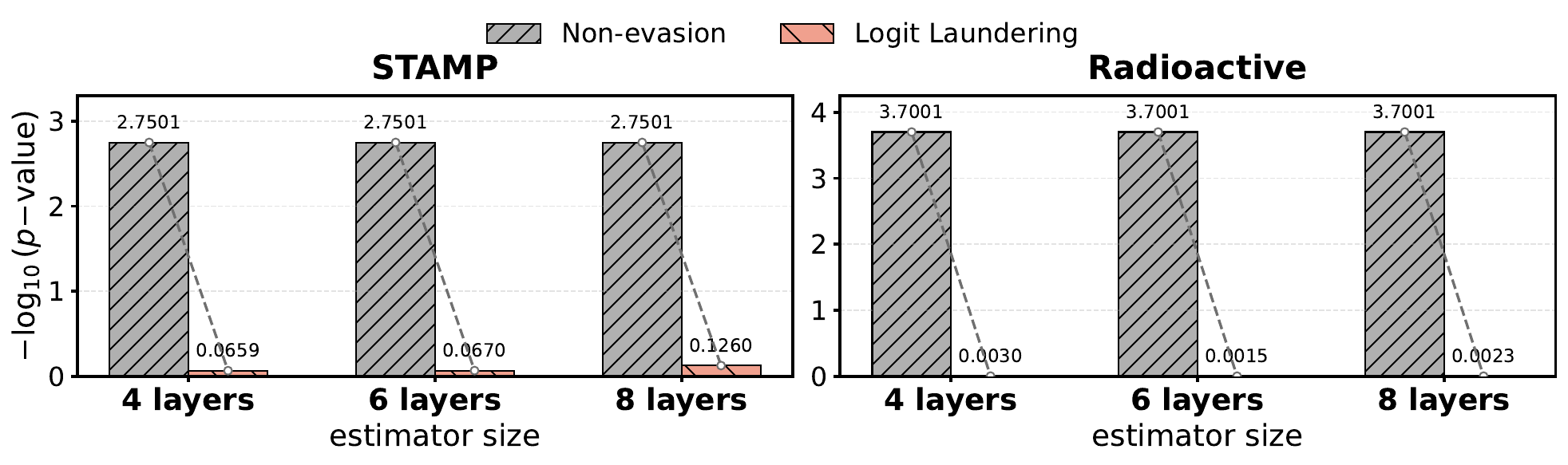}
    \caption{Detectable signals under different estimator sizes}
    \label{fig:layers_neglog10_bar}
      \vspace{-0.5em}
\end{figure}

\vspace{0.2em}
\noindent\textbf{Effect of Statistical Bias Estimator Size}.
Figure~\ref{fig:layers_neglog10_bar} indicates that the size of the estimator (i.e., the number of Transformer layers used to construct it) has little impact on evasion performance. Even with a smaller scale (e.g., using only a few shallow layers), our framework can still achieve satisfactory evasion performance. That is, Logit Laundering is insensitive to estimator size: even a small estimator is effective, allowing its size to be adjusted to the available computational budget with little impact on evasion performance.

 \begin{figure}[t]
    \centering
    \subfloat[\small Segment ratio $\alpha$.]{
        \includegraphics[width=0.39\linewidth]{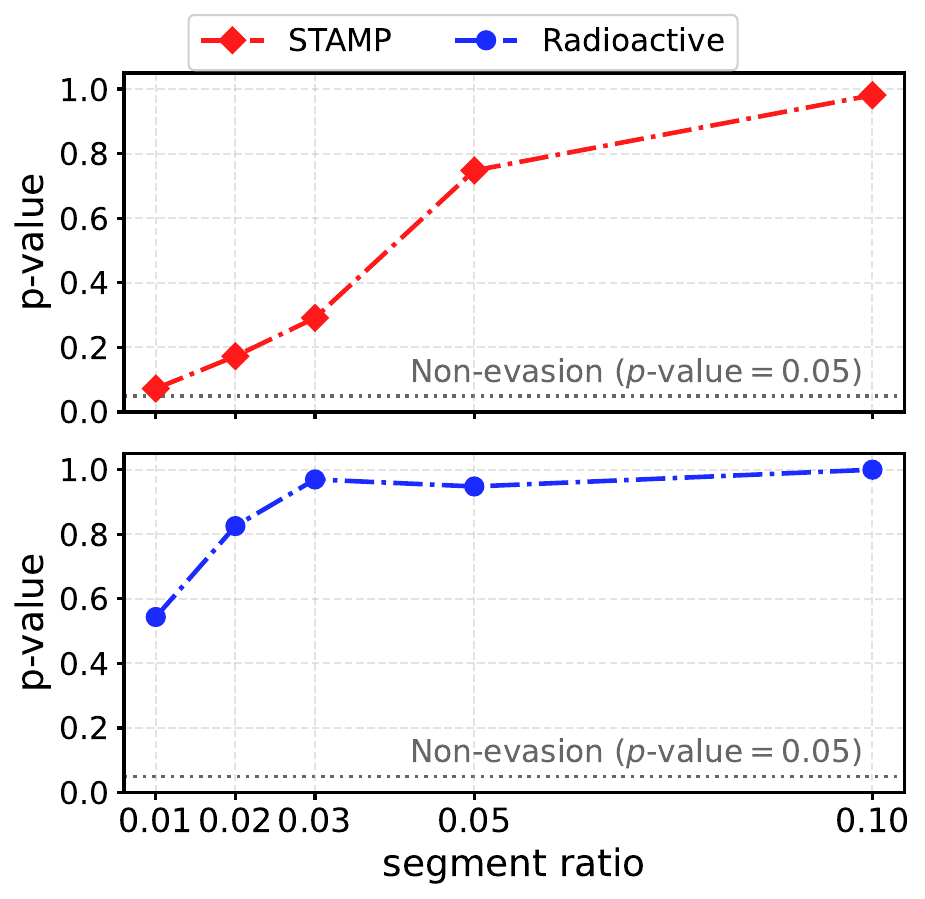}
    }
    \hfill
    \subfloat[\small Smoothing parameter $S$.]{
        \includegraphics[width=0.57\linewidth]{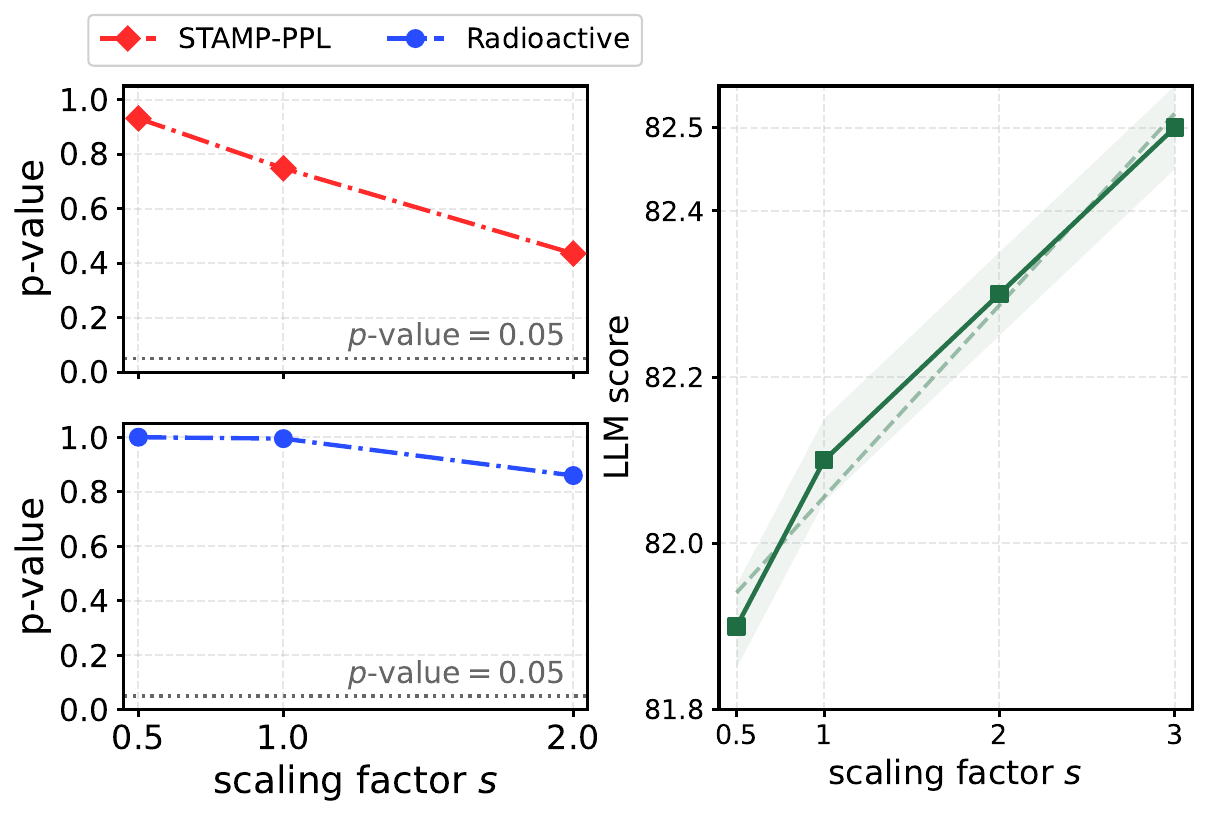}
    }
\caption{Hyperparameter analysis of segment ratio $\alpha$ and scaling smoothing parameter $S$.}
    \label{fig:hy_alpha}
  \vspace{-0.5em}
\end{figure}

\vspace{0.3em}
\noindent\textbf{Sensitivity to segment ratio $\alpha$ and scaling parameter $S$}. As shown in Figure~\ref{fig:hy_alpha} (a), increasing the segment ratio $\alpha$ consistently reduces auditing detectability, indicating that longer segments lead to stronger suppression and improved evasion performance. However, excessively large $\alpha$ may introduce broader perturbations to the prediction distribution, thereby interfering with knowledge expression in generated outputs. As shown in Figure~\ref{fig:hy_alpha} (b), as $S$ increases, the p-values of both detection methods slightly decrease, while remaining at strong evasion performance. This indicates that the evasion effect is stable across different choices of $S$. The possible reason is that a larger $S$ makes the adaptive suppression smoother and less aggressive, slightly reducing its ability to weaken watermark signals while better preserving the base model distribution and minimizing model generation degradation.

\vspace{0.3em}
\noindent\textbf{Effect of confidence scaling.} Table~\ref{tab:confidence_scaling_ablation} shows that removing confidence scaling and replacing the adaptive coefficient with a fixed $\lambda$ substantially degrades generation quality: the MAUVE score drops sharply, indicating that the generated texts are not natural. That is, confidence-aware scaling is necessary for preserving more natural outputs while still weakening the watermark signal effectively. The $\lambda$ ranges are computed under a fixed JS divergence budget of $[0.01, 0.1]$ (as Sec.~\ref{method: scaling} described), and thus vary across models. See Appendix~\ref{app: results} for more ablations.

\begin{figure}[t]
    \centering
    \includegraphics[width=0.97\columnwidth]{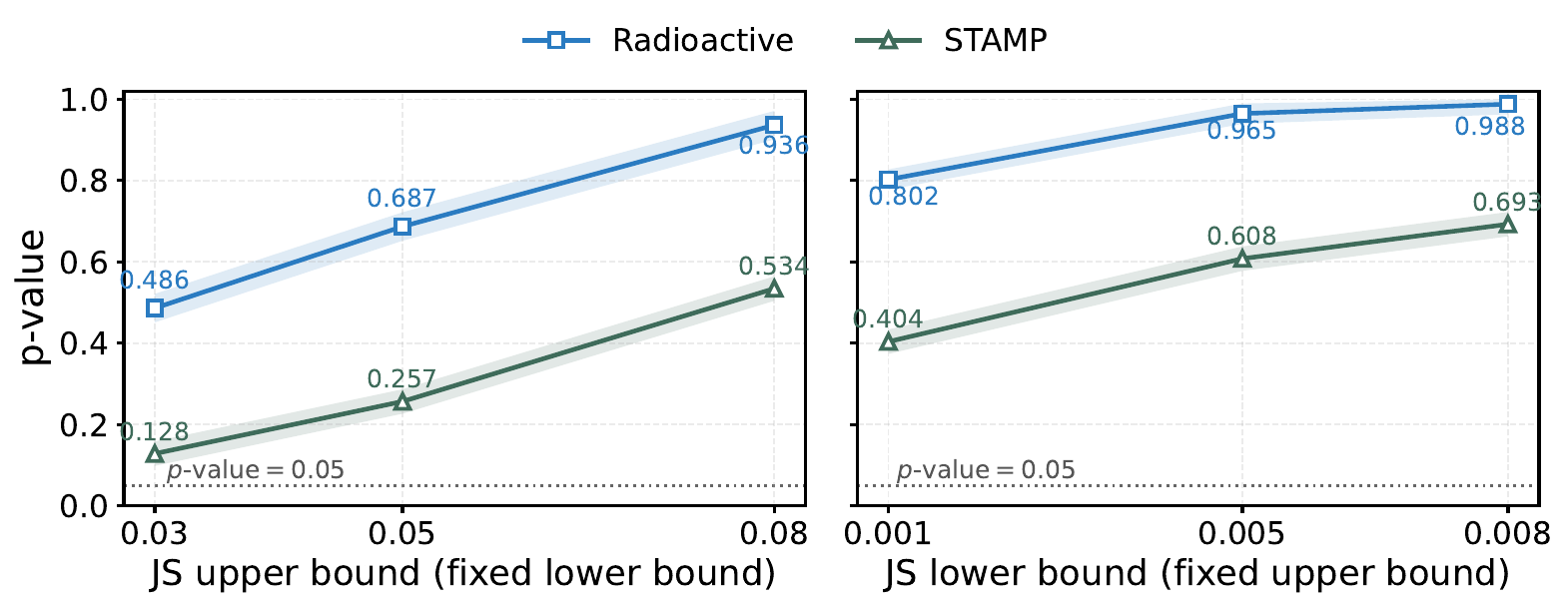}
    \caption{Evasion performance under different JS budget range settings, $[\tau_{\min}, \tau_{\max}]$. }
    \label{fig:js_budget_two_panels}
\end{figure}

\vspace{0.3em}
\noindent\textbf{JS budget range.} The JS budget range controls the allowable shift of the reshaped distribution from the base model distribution. Figure~\ref{fig:js_budget_two_panels} shows that increasing the JS upper/lower bound consistently raises the p-values of both STAMP and Radioactive, indicating that a more relaxed budget enables stronger logit reshaping and more effectively weakens watermark signals. Conversely, a smaller upper bound can still maintain effective evasion, but it limits the degree of signal suppression and induces a smaller distributional shift, which better preserves model utility and knowledge expression. The budget range can be adjusted according to specific requirements by non-compliant LLM providers, while the pre-defined range adopted in the main evaluation offers a favorable trade-off for most LLMs.

\begin{table}[t]
  \caption{Ablation of confidence scaling module}
    \vspace{-0.5em}
  \centering
  \small
  \setlength{\tabcolsep}{5pt}
  \renewcommand{\arraystretch}{0.7}
  \resizebox{0.85 \columnwidth}{!}{%
  \begin{tabular}{l l l c c}
    \toprule
    \textbf{Model} & \textbf{Variant} & \textbf{$\lambda$ sched.} & \textbf{MAUVE $\uparrow$} & \textbf{$p$-value $\uparrow$} \\
    \midrule

    \multirow{4}{*}{Llama-3.2-1B}
      & adaptive
      & $[0.170, 0.560]$
      & 0.107 {\scriptsize\phantom{$\downarrow\downarrow$}}
      & 0.748 \\
      & w/o scaling
      & $\lambda=0.170$
      & 0.018 {\scriptsize\textcolor{gray!70}{$\downarrow\downarrow$}}
      & 0.044 \\
      & w/o scaling
      & $\lambda=0.365$
      & 0.016 {\scriptsize\textcolor{gray!70}{$\downarrow\downarrow$}}
      & 0.752 \\
      & w/o scaling
      & $\lambda=0.560$
      & 0.018 {\scriptsize\textcolor{gray!70}{$\downarrow\downarrow$}}
      & 0.982 \\
    \midrule

    \multirow{4}{*}{Qwen2.5-1.5B}
      & adaptive
      & $[0.140, 0.450]$
      & 0.151 {\scriptsize\phantom{$\downarrow\downarrow$}}
      & 0.412 \\
      & w/o scaling
      & $\lambda=0.140$
      & 0.019 {\scriptsize\textcolor{gray!70}{$\downarrow\downarrow$}}
      & 0.069 \\
      & w/o scaling
      & $\lambda=0.295$
      & 0.017 {\scriptsize\textcolor{gray!70}{$\downarrow\downarrow$}}
      & 0.445 \\
      & w/o scaling
      & $\lambda=0.450$
      & 0.017 {\scriptsize\textcolor{gray!70}{$\downarrow\downarrow$}}
      & 0.935 \\
    \bottomrule
  \end{tabular}%
  }
  \vspace{-0.5em}
  \label{tab:confidence_scaling_ablation}
\end{table}

 \subsection{Implications for Defenses}
Our results suggest that robust data-use auditing requires more than stronger watermark design; it also requires a more robust detection pipeline. Existing watermark-based auditing methods often assume that the LLM provider-exposed outputs, logits, or token-level statistics faithfully reflect the model's learned knowledge and use of the audited data. However, this assumption can be violated when a non-compliant provider manipulates the inference-time output distribution, leading to false negatives (FNs). Therefore, auditing pipelines should include a post-detection analysis step to identify suspicious FNs caused by suppression of detectable signals. 

To this end, we propose an exploratory defensive approach, termed \textit{True-key Placebo Check}. It compares the model’s green-token ratio under the real watermark key with the ratios under random placebo keys, and checks whether an anomalous shift appears specifically under the real key. For a clean model, the real and random keys should yield similar ratios in expectation, since no key-specific signal should be present. In contrast, if logit laundering suppresses signals associated with the real key, the real-key ratio may shift in the opposite direction and fall below the random-key baseline. This true-key shift is then used to flag suspicious FNs for further forensic auditing of the provider’s decoding process.

We conduct a small-scale pilot study on one book. For each prefix, we take the model's generated token and check whether it is green under the real watermark key. We then repeat the same computation using 10 random placebo keys and report the average green token frequency ratio under these placebo keys (more experimental details are provided in Appendix~\ref{app: defense}). True-key shift is defined as the difference between the real-key green ratio and the random-key mean. 
As shown in Table~\ref{tab:random-key-suppression}, the clean model that has not been trained on unauthorized copyrighted data shows a near-zero shift; the non-evasion model, which used copyright data for training, exhibits a positive shift, but the laundered one shifts in the opposite direction, with its true-key green ratio falling below the random-key baseline. It suggests that the LLM provider may actively suppress the output signals associated with the real key, providing a potential post-detection indicator for suspicious false negatives. This reverse shift can therefore serve as a basis for auditors to request additional forensic information from the provider, such as more details about the decoding and inference-time process.

\begin{table}[t]
\caption{
True-key placebo check.
A negative shift indicates signal suppression and may flag a suspicious false negative.
}
\centering
\small
\renewcommand{\arraystretch}{0.8}
\resizebox{0.8\columnwidth}{!}{
\begin{tabular}{lccc}
\toprule
Model & True-key ratio & Random-key mean & Shift \\
\midrule
Clean & 0.2496 & 0.2476 & +0.0019 \\
Non-evasion & 0.2781 & 0.2503 & +0.0278 \\
Laundered & 0.2400 & 0.2513 & -0.0112 {\tiny \textcolor{red}{\textbf{FN flag}}} \\
\bottomrule
\end{tabular}
}
\label{tab:random-key-suppression}
\end{table}

\section{Conclusion}
In this paper, we present a systematic adversarial study of watermark-based data-use auditing in LLMs. We show that straightforward data-level evasion can distort the training corpus while leaving residual detectable signals, motivating our focus on model-level evasion. We then propose Logit Laundering, an evasion framework that reshapes output logits to suppress detectable statistical signals without modifying model parameters. By integrating a statistical bias estimator with confidence-aware scaling, Logit Laundering weakens watermark evidence while largely preserving newly acquired copyrighted knowledge and general model utility. Extensive evaluations across five foundation LLMs and multiple state-of-the-art auditing methods demonstrate that existing auditing pipelines can be vulnerable to inference-time output manipulation. Our findings highlight the need to improve the robustness of data-use auditing pipelines against such evasion. Our future work would jointly explore stealthier evasion strategies and defensive detection mechanisms to strengthen data-use auditing robustness.

\section*{Acknowledgement}

This work is supported by the National Natural Science Foundation of China (No. 62372057 and No. 6240073908), BUPT-China Unicom Joint Innovation Center (2025-STHZ-BJYDDX-010), and the Fundamental Research Funds for the Central Universities (No. 500422828), and partially supported by BUPT Kunpeng \& Ascend Center of Cultivation.

\section*{Ethical Considerations}
This paper studies the robustness of watermark-based data-use auditing for LLMs under non-compliant provider behavior, focusing on how inference-time output manipulation can weaken detectable statistical signals. Our goal is to reveal limitations of current auditing pipelines, not to encourage data misuse, copyright infringement, or audit circumvention. We analyze data-level and model-level evasion and discuss post-detection checks for suspicious false negatives and signal suppression. Our results suggest that missing watermark signals should not be treated as conclusive evidence of non-use under suspicious evasion behavior. All experiments are for academic analysis, use no private user data, and rely on public, licensed, or research-prepared datasets.

{\footnotesize \bibliographystyle{acm}
\bibliography{sample}}
\clearpage
\appendix

\section{Data-level Evasion Baselines}  \label{app: baselines}
In this section, we provide details about the data-level evasion baseline methods.
While this work focuses on evasion against watermark-based data-use auditing in LLMs, we additionally adapt several representative baselines from related research to contextualize our contributions within existing literature. In particular, we incorporate attack methods from the LLM watermarking literature, which aim to directly disrupt the statistical biases present in text generated by watermarked LLMs~\cite{Zhang2025CharacterLevelPD, he2024watermarks,liang2025watermarkunderfire,paraattack}, which include:
\begin{itemize}
	\item \textbf{Char (homoglyph)}~\cite{Zhang2025CharacterLevelPD}: It perturbs the generated text by replacing selected characters with visually similar homoglyphs. For instance, the character `a' can be replaced with `\`{a}' (U+00E0). Such homoglyph substitution disrupts the tokenization process, frequently causing a single word to be segmented into multiple subword tokens, thereby enlarging the effective attack surface for interfering with watermark signals.	
	\item \textbf{Char (insertion)}~\cite{Zhang2025CharacterLevelPD}: It inserts zero-width characters (\emph{e.g.}, U+200B) or spaces at specific positions in the text. While these inserted characters do not affect visual representation, they disturb the token split of the original text, disrupting the statistical features of the watermark.

	\item \textbf{Char (typo)}~\cite{Zhang2025CharacterLevelPD}: It introduces spelling errors by replacing characters with adjacent ones on the keyboard (\emph{e.g.,} replacing `t' with `r'). Such a disturbance can disrupt the character sequence of the original text, affecting the statistical features of the watermark.
	
	\item \textbf{Token Swap}~\cite{liang2025watermarkunderfire}: It disturbs the text by randomly swapping the positions of tokens in the sentence. This disturbance changes the order structure of the text and can effectively disrupt watermark signals.
	
	\item \textbf{Token Synonym}~\cite{liang2025watermarkunderfire}: it attack replaces tokens with their synonyms using a predefined lexical database (\emph{e.g.,}, WordNet~\cite{wordnet}), while maintaining semantic consistency of the text. By altering the use of vocabulary, this attack disrupts token-based watermark information.
	
	\item \textbf{Paraphrase}~\cite{paraattack}: it rewrites sentences using a paraphrase generation model, producing semantically equivalent sentences with different syntactic structures. This method can significantly change the form of the text, potentially bypassing some watermark detection.
	
	\item \textbf{Translation}~\cite{he2024watermarks}: it changes the surface structure of the watermark text by translating it into another language and then translating it back into the original language. Translation attacks introduce subtle semantic changes through cross-language transformations, thereby disrupting watermark information.
	
\end{itemize}

\begin{figure}[t]
    \centering
    \includegraphics[width=0.9 \columnwidth]{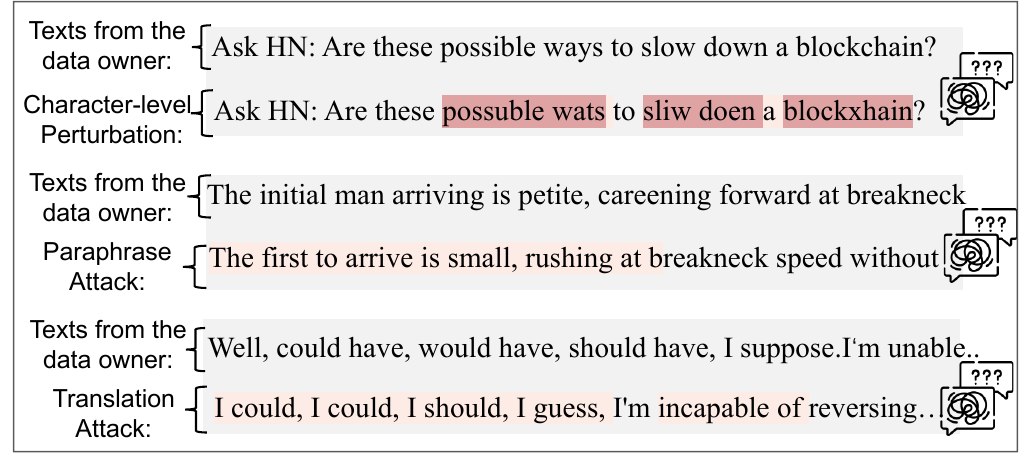}
    \caption{
Text transformations~\cite{Zhang2025CharacterLevelPD, paraattack} would distort the original content, impairing their value as training data. 
    }
    \label{fig:example_text_level}
\end{figure}

These baselines are originally designed to attack watermarked generated text, where preserving the training value of the transformed text is not the primary goal. In our data-use auditing scenario, however, the transformed texts are still expected to serve as training data. As illustrated in Figure~\ref{fig:example_text_level}, such transformations can introduce typos, alter phrasing, or change semantic nuances, thereby distorting the original copyrighted content and reducing its value for model training.

\section{Confidence-aware Scaling Analysis}  \label{app: scaling-analysis}

This appendix provides additional analysis for the confidence-aware scaling rule used in Logit Laundering. The goal is to explain why the top-1/top-2 logit margin can serve as a practical confidence signal for determining how aggressively the output logits can be reshaped at each decoding step. The top-1/top-2 logit margin can reflect next-token confidence~\cite{yoshikawa-okazaki-2023-selective}: a larger margin means that the model is more confident in its top prediction, while a smaller margin indicates greater uncertainty among competing candidates.

\noindent\textbf{Stability intuition.}
Let $m_t = z_b^{(1)} - z_b^{(2)}$ denote the margin between the largest and second-largest base-model logits at decoding step $t$. A larger margin indicates that the dominant candidate is well separated from its closest competitor. Under an additive logit perturbation $\Delta_t$, if $\|\Delta_t\|_\infty < m_t/2$, the identity of the top-1 token is guaranteed to remain unchanged. Therefore, positions with larger margins provide a larger robustness buffer against logit reshaping. In contrast, small-margin positions have flatter candidate structures, making token rankings and sampling outcomes more sensitive to perturbations.

\noindent\textbf{Empirical perturbation study.}
To empirically examine this sensitivity, we conduct a small-scale perturbation experiment on next-token distributions. For each selected decoding position, we apply a small Gaussian perturbation to the first-step logits:
\begin{equation}
z'_t = z_t + \epsilon,
\qquad
\epsilon \sim \mathcal{N}(0,\sigma^2 I),
\end{equation}
where $\sigma=0.1$. The perturbation is applied only at the first decoding step. The subsequent continuation is generated using the same decoding strategy as the original branch. We evaluate three decoding strategies: greedy decoding, top-$k$ sampling with $k=40$, and nucleus sampling with $p=0.9$, all using temperature $0.7$.

We group decoding positions by the top-1/top-2 logit margin $m_t$. Large-margin positions correspond to more concentrated predictions, while small-margin positions correspond to flatter candidate structures with smaller effective decision margins. We report three metrics: \textit{Candidate Overlap}, the Jaccard overlap between the original and perturbed decoding candidate sets; \textit{Distribution Shift from Base}, the probability mass that moves outside the original base model candidate set after perturbation; and \textit{Continuation Drift}, a continuation deviation score computed as one minus the embedding cosine similarity between the original and perturbed continuations. Higher Distribution Shift and Continuation Drift indicate stronger sensitivity, while higher Candidate Overlap indicates stronger distributional stability.

As shown in Table~\ref{tab:confidence-sensitivity}, small-margin positions exhibit weaker candidate-set stability, larger distributional shift from the base model, and larger continuation-level drift under the same perturbation scale. The effect is most pronounced under top-$k$ sampling, where small-margin positions show both lower candidate overlap and larger continuation drift. Under top-$p$ sampling, the distributional shift is smaller, but the candidate overlap and continuation drift still indicate greater instability for small-margin positions. These observations support the design of margin-based adaptive scaling: confident positions with larger top-1/top-2 margins can tolerate stronger reshaping, while uncertain positions require more conservative suppression to avoid unnecessary generation degradation.

\begin{table}[H]
\centering
\small
\caption{Sensitivity to first-step logit perturbations across margin groups.}
\label{tab:confidence-sensitivity}
\setlength{\tabcolsep}{5pt}
\renewcommand{\arraystretch}{1.05}
\resizebox{\columnwidth}{!}{
\begin{tabular}{l|l|ccc}
\toprule
Decoding & Margin Group & Candidate Overlap & Dist. Shift & Cont. Drift \\
\midrule
Greedy & Large & 1.000 & 0.136 & 0.000 \\
Greedy & Small & 0.917 & 0.748 & 0.066 \\
\midrule
Top-$k$ & Large & 0.952 & 0.000 & 0.056 \\
Top-$k$ & Small & 0.928 & 0.006 & 0.130 \\
\midrule
Top-$p$ & Large & 1.000 & 0.000 & 0.059 \\
Top-$p$ & Small & 0.895 & 0.005 & 0.070 \\
\bottomrule
\end{tabular}}
\end{table}

\section{Additional Experimental Setting Results} \label{app: setting}

\subsection{Dataset Watermarking Details} \label{app: dataset}
Following~\cite{jovanović2025wardprovableragdataset}, we generate watermarked datasets using two instruction-tuned LLMs: Llama-3-8B-Instruct~\cite{grattafiori2024llama3herdmodels}, Qwen-2.5-14B-Instruct~\cite{qwen2.5}, Gemma-2-9B-IT~\cite{gemmateam2024gemmaopenmodelsbased} and Stable-LM-2-12B-Chat~\cite{bellagente2024stable}. All LLMs are instrumented with a \textbf{red-green watermarking scheme} during text generation and paraphrasing. The watermarking paraphrase prompt is:
\texttt{"You are a rewriting model. Rewrite the following text clearly and naturally, keeping the same meaning and about the same length. Do not add, remove, or comment on any information. Do not include phrases like 'Here is the paraphrased version' or any explanation. Output ONLY the rewritten text itself, nothing else.\textbackslash n\textbackslash n\#\#\# Original Text:\textbackslash n"}

Specifically, we adopt watermarking hyperparameters $\delta = 4.0$ and $\gamma = 0.25$, where $\gamma$ denotes the fraction of tokens assigned to the green list, and $\delta$ specifies the logit bias added to green-list tokens at each generation step. We further set the watermark window size to $w=1$, meaning that the watermark bias is applied independently at each token position without cross-token dependency. These settings follow standard configurations in prior watermarking studies and provide a strong watermark signal for downstream auditing and adversarial evaluations. 

\subsection{Baseline Implementation Details}
In this section, we provide detailed implementation settings for the baseline in our main results.
For the data-level transformations, we adopt a unified edit rate of $0.3$, which controls the maximum proportion of characters or tokens that can be modified in each attack. For character-level attacks, including Char (Homoglyph), Char (Insertion), and Char (Typo), we follow the settings used in prior work. Specifically, Char (Homoglyph) relies on a predefined list of homoglyph pairs, each consisting of a character and its visually similar counterpart. The attack randomly selects a character in the text and replaces it with its corresponding homoglyph if such a pair exists. Char (Insertion) randomly selects a position in the text and inserts a character sampled from a predefined character set, including letters, numbers, and symbols. Char (Typo) uses a predefined set of common keyboard errors, where a randomly selected character is replaced with an adjacent key on the keyboard. The keyboard error range is set to $\texttt{typo\_range}=1$, meaning that each character is only replaced with one of its immediate neighbors. For the Token Swap attack, the edit rate determines the probability of triggering an attack for each token. Once triggered, two words in the sentence are randomly swapped to disrupt the syntactic structure. For the Token Synonym attack, only words that have available synonyms in WordNet are considered for replacement, with a focus on nouns, verbs, and adjectives. The replacement probability is set to $\texttt{synonym\_p}=0.5$, meaning that each eligible word has a 50\% chance of being replaced by one of its synonyms. The Paraphrase attack employs the DIPPER~\cite{krishna2023paraphrasing} model to generate paraphrased text, with parameters set to $\texttt{lex}=20$ (vocabulary diversity), $\texttt{order}=0$ (order diversity), and $\texttt{sent\_interval}=3$ (sentence interval), producing semantically equivalent but structurally different sentences. Finally, the Translation attack applies a back-translation strategy (English $\rightarrow$ French $\rightarrow$ English) using the \texttt{seamless-m4t-v2-large}~\cite{barrault2023seamlessm4t} model to generate diverse phrasings and alter surface expressions through translation variability.

\begin{table}[t]
\caption{
Release time of evaluated foundation models and publication time of our collected copyrighted book corpus.
}
\centering
\small
\begin{tabular}{lc}
\toprule
Target LLM & Release time \\
\midrule
LLAMA-3.2-1B-Instruct & 2024-09 \\
LLAMA-3.2-3B-Instruct & 2024-09 \\
Qwen-2.5-1.5B-Instruct & 2024-09 \\
Gemma-2-2B-IT & 2024-07 \\
Stable-LM-2-1.6B-Chat & 2024-02 \\
\midrule
Our copyrighted-book corpus & 2025 \\
\bottomrule
\end{tabular}

\label{tab:model_release_time}
\end{table}

\subsection{Copyright Corpus and MCQ Set Construction} \label{app: mcq-construction}
As shown in Table~\ref{tab:model_release_time}, all evaluated foundation models were released before 2025, while our copyrighted-book corpus consists of books published in 2025. This temporal separation reduces potential overlap with the models' original pretraining data, providing a cleaner setting for evaluating unauthorized continual pretraining and subsequent auditing evasion.It should be noted that this newly collected corpus is sufficient for our evaluation, because our goal is not to compare performance across multiple datasets, but to measure how watermark signals change before and after evasion within the same watermarking, continual pretraining, and auditing pipeline. Accordingly, the generality of our method is mainly reflected in its consistent effectiveness across different auditing mechanisms and multiple foundation models.

\noindent\textbf{Corpus Construction.} We construct a copyright corpus~$\mathbb{D}_{\mathrm{wm}}$ from 50 books published in 2025, minimising overlap with the foundation models' pretraining data.
The books span contemporary commercial fiction genres including romance, fantasy, thriller/mystery, science fiction, and literary fiction.
Each book undergoes text cleaning to remove non-narrative elements (table of contents, copyright pages, headers/footers, etc.), after which the body text is segmented into paragraph-level samples.
Paragraphs are capped at 512~tokens; passages exceeding this limit are split at natural boundaries to preserve semantic integrity.
We then perform evenly spaced index-based sampling to select approximately 100 non-overlapping paragraphs per book, ensuring coverage of the beginning, middle, and end of each work; books with fewer than 100 paragraphs are included in full.
This procedure yields 4,959 paragraphs in total.

\vspace{0.3em}
\noindent\textbf{MCQ Set Construction.}
We construct a multiple-choice question-answering (MCQ) benchmark to evaluate whether a model acquires specific factual knowledge from the copyrighted corpus after continual pretraining. Claude-Sonnet-4.6 and GPT-5.4 are each used to independently generate 1,000 questions from passages in~$\mathbb{D}_{\mathrm{wm}}$, with questions roughly evenly distributed across the 50 books ($\sim$20 questions per book), yielding two disjoint MCQ subsets. Each question consists of one correct answer and three plausible distractors. During generation, models are instructed to base questions solely on the supplied passage and avoid using external knowledge. Questions are designed to target specific facts, events, character relationships, concept definitions, or causal relationships stated in the passage, rather than broad thematic judgments answerable without close reading. A sampled subset of generated items undergoes human quality review. Annotators verify that: 
\begin{enumerate}[label=(\roman*), leftmargin=2.5em, labelwidth=2em, align=left, itemsep=0.2em]
    \item the stem and correct answer are directly supported by the passage;
    \item distractors are plausible but unambiguously incorrect;
    \item items cannot be answered using common sense or surface-form cues alone; and
    \item no duplicates, ambiguities, or items with multiple defensible answers are present.
\end{enumerate}
Items failing these criteria are removed or revised.

\subsection{General Utility Evaluation Details} \label{app: general-utility}
\noindent\textbf{Generation-Quality Evaluation.}
To assess the impact of Logit Laundering on general text generation quality, we curate a held-out evaluation set of 500 samples drawn from seven heterogeneous sources, as detailed in Table~\ref{tab:general_eval_sources}.
Samples fall into two categories: \emph{out-of-training sources}---Wikipedia~(en), Books3, and StackExchange, drawn from the Pile dataset but excluded from training corpus~$\mathbb{D}$---and \emph{in-training sources}---Copyrighted Books, Pile-CC, Hacker News, and arXiv---which are part of~$\mathbb{D}$.
After filtering for quality (removing malformed, code-heavy, repetitive, or noisy samples), the final set has a mean length of 209.4~tokens and a median of 210~tokens.

For each sample, the first 100~tokens serve as the prompt prefix and the remainder as the human reference continuation.
Each evaluated model autoregressively generates a continuation of the same length as the reference, conditioned on the identical prompt, ensuring a fair length-normalised comparison across all metrics.
Generation parameters follow standard open-ended text generation practice: we set temperature $0.7$ and apply nucleus (top-$p$) sampling with $p = 0.9$.
No system prompt or instruction template is used; the model receives the raw prefix as a plain continuation task.
All models are evaluated under identical decoding settings to ensure comparability.

\begin{table}[h]
\centering
\caption{Statistics of the 500 generation-quality evaluation samples,
grouped by whether the source belongs to the continual pretraining
corpus~$\mathbb{D}$.}
\label{tab:general_eval_sources}
\footnotesize
\setlength{\tabcolsep}{5pt}
\renewcommand{\arraystretch}{0.95}
\begin{tabular}{lrrr}
\toprule
\textbf{Source} & \boldmath{$n$} & \textbf{Mean (tokens)} & \textbf{Median (tokens)} \\
\midrule
\rowcolor{gray!12}
\multicolumn{4}{l}{\textit{Out-of-training sources}} \\
Wikipedia (en) &  84 & 217.6 & 220 \\
Books3         &  83 & 219.9 & 222 \\
StackExchange  &  83 & 217.0 & 219 \\
\rowcolor{gray!12}
\multicolumn{4}{l}{\textit{In-training sources}} \\
Copyrighted Books &  63 & 235.1 & 229 \\
Pile-CC           &  63 & 192.2 & 200 \\
Hacker News       &  62 & 189.2 & 191 \\
arXiv             &  62 & 185.7 & 186 \\
\midrule
\textbf{Overall} & \textbf{500} & \textbf{209.4} & \textbf{210} \\
\bottomrule
\end{tabular}
\end{table}

\vspace{0.3em}
We evaluate generation quality using three types of metrics.
\textbf{Perplexity (PPL)} measures local fluency by computing the perplexity of model-generated continuations under GPT-2-XL; lower is better.
\textbf{MAUVE} measures the distributional similarity between model-generated and human-written text in an embedding space using a GPT-2-based feature extractor; higher scores reflect closer alignment with human writing.
\textbf{LLM-Judge} scores employs GPT-5.4 and Claude-Sonnet-4.6 as independent judges to directly assess the quality of each model-generated continuation.
For each sample, the judge is provided with the prompt prefix and the corresponding generated continuation, and is asked to rate the overall quality of the continuation on a 0--100 scale along dimensions such as fluency, coherence, and informativeness.

\vspace{0.5em}
\noindent\textbf{Standard Downstream Task Evaluation.}
We additionally evaluate general downstream utility using \texttt{lm-evaluation-harness} on ARC-Challenge, MMLU, TriviaQA, and GSM8K. Due to computational cost, we use 5-shot evaluation and evaluate 100 samples for each task.
For MMLU, the 100-sample limit is applied separately to each subject subtask before aggregating the results.
Following~\cite{eval-harness}, for all generation-based tasks, sampling is disabled (\texttt{do\_sample=false}) and the temperature is set to 0.0.

\begin{table}[t]
\centering
\caption{Recommended $\lambda$ under each $\tau$ value for all five foundation models.}
\label{tab:alpha_max}
\footnotesize
\setlength{\tabcolsep}{7pt}
\renewcommand{\arraystretch}{0.75}
\begin{tabular}{cccccc}
\toprule
\textbf{$\tau_{\min}$} \& \textbf{$\tau_{\max}$} & \textbf{LA-1B} & \textbf{Qw-1.5B} & \textbf{LA-3B} & \textbf{Ge-2B} & \textbf{SLM-1.6B} \\
\midrule
\rowcolor{gray!10}
0.01 & 0.17 & 0.14 & 0.15 & 0.14 & 0.12 \\
0.02 & 0.24 & 0.19 & 0.21 & 0.20 & 0.18 \\
0.03 & 0.29 & 0.24 & 0.26 & 0.25 & 0.22 \\
0.04 & 0.34 & 0.28 & 0.30 & 0.28 & 0.25 \\
0.05 & 0.38 & 0.31 & 0.34 & 0.32 & 0.28 \\
0.06 & 0.42 & 0.34 & 0.37 & 0.35 & 0.31 \\
0.07 & 0.46 & 0.37 & 0.40 & 0.38 & 0.34 \\
0.08 & 0.49 & 0.40 & 0.43 & 0.41 & 0.36 \\
0.09 & 0.53 & 0.43 & 0.47 & 0.43 & 0.39 \\
\rowcolor{gray!10}
0.10 & 0.56 & 0.45 & 0.49 & 0.46 & 0.41 \\
\bottomrule
\end{tabular}
\end{table}

\vspace{0.3em}
\noindent\textbf{JS Budget Range.} Table~\ref{tab:alpha_max} reports the recommended $\lambda_{min}$ and $\lambda_{max}$ for all five foundation models across JS budgets ranging from 0.01 to 0.10.  As the JS budget increases, $\tau_{\max}$ grows monotonically for every model, reflecting that a more relaxed distributional constraint permits stronger logit suppression. The range $[\tau_{\min}, \tau_{\max}] = [0.01, 0.1]$ adopted in the main evaluation offers a balanced compromise across most models.

\subsection{Experimental Details of Defenses}
\label{app: defense}
We conduct the True-key Placebo Check on 10 paragraphs from one book. In this section, we use the same watermark configuration as in the main experiments: $\gamma$ = 0.25, $\delta$= 3.5, and window size w = 1. For the same prefix, different keys induce different green/red partitions of the vocabulary while keeping the same greenlist fraction gamma. The real key is the private watermark key used to generate the watermarked data, whereas the placebo keys are randomly sampled keys that were never used in the watermarking process. At each token position, we first obtain the model's top-1 next-token prediction. We then evaluate the same predicted token under multiple key-induced green/red partitions: one induced by the real private watermark key and others induced by randomly sampled placebo keys. We evaluate three models: a clean foundation model that has not been trained on the audited watermarked data, a non-evasion model that has been trained on unauthorized copyrighted data without applying any audit-evasion strategy, and a logit-laundering model that applies our inference-time reshaping framework to evade detection. As shown in Table~\ref{tab:random-key-suppression} , we report the real-key green-token ratio, the average green-token ratio under the placebo keys, and the suppression shift, defined as the difference between the two. Positive shifts indicate residual preference toward the real watermark key, while negative shifts suggest potential key-specific suppression. 

\section{Additional Experimental Results}  \label{app: results}

\begin{algorithm}[t]
\caption{Logit Reshaping Algorithm}
\small
\label{alg:reshaping}
\textbf{In:} Public data $D_{\mathrm{pub}}$; watermarked data $D_{\mathrm{wm}}$;
base model $\Theta_b$; statistical bias estimator $\Theta_c$; flatten weight $\eta$;
control weight $\lambda$; decoding length $T$.\\
\textbf{Out:} Reshaped generation $\{x_t\}_{t=1}^{T}$.

\begin{algorithmic}[1]
\REPEAT
    \STATE Sample $x \sim D_{\mathrm{pub}}$, $(x_{ws},y)\sim D_{\mathrm{wm}}$
    \STATE $L_{\mathrm{flat}} \leftarrow D_{\mathrm{KL}}(U \,\|\, P_{\Theta_c}(\cdot|x))$;  $L_{\mathrm{fit}} \leftarrow -\log P_{\Theta_c}(y|x_{ws})$
    \STATE Update $\Theta_c$ by minimizing $L_c = L_{\mathrm{fit}} + \eta L_{\mathrm{flat}}$
\UNTIL{convergence}

\STATE $\mathbf{x}\leftarrow[\ ]$ \textcolor{gray}{\textit{// Inference-time Logit Reshaping}}
\FOR{$t=1,2,\ldots,T$}
    \STATE $z_b \leftarrow \mathrm{Logits}(\Theta_b \mid \mathbf{x}_{<t})$;  $z_c \leftarrow \mathrm{Logits}(\Theta_c \mid \mathbf{x}_{<t})$
    \STATE $z \leftarrow z_b - \lambda_t \cdot z_c$
    \STATE $x_t \leftarrow \mathrm{Decode}(z)$
    \STATE $\mathbf{x} \leftarrow \mathbf{x} \Vert x_t$
\ENDFOR
\STATE \textbf{return} $\mathbf{x}$
\end{algorithmic}
\end{algorithm}

\begin{table}[t]
\caption{Knowledge Retention Performance on the Claude Sonnet 4.6-generated MCQ set.}
  \centering
  \setlength{\tabcolsep}{4pt}
  \renewcommand{\arraystretch}{1}
  \resizebox{\columnwidth}{!}{%
  \begin{tabular}{l|c|c|c|c|c}
    \toprule
Copyright Knowledge &
LA-1B &
LA-3B &
Qw-1.5B &
Ge-2B &
SLM-1.6B\\
    \cmidrule(lr){1-6}
    \rowcolor{gray!15}
Foundation Model &
    33.8\% & 80.7\% & 87.1\% & 76.2\% & 29.4\% \\

Non Evasion &
    86.6\% & 93.8\% & 94.4\% & 94.1\% & 65.7\% \\
     \hline
    \multicolumn{6}{c}{\textit{Data-level Evasion}} \\
    \addlinespace[2pt]
    \hline
    
Char (homoglyph)  &
    44.6\% & 87.7\% & 87.2\% & 58.2\% & 39.3\% \\
Char (insertion) &
    44.3\% & 88.6\% & 88.2\% & 60.1\% & 40.8\% \\
Char (typo) &
    44.3\% & 88.3\% & 88.0\% & 59.6\% & 38.8\% \\
Token Swap&
    53.0\% & 87.6\% & 88.5\% & 68.6\% & 42.3\% \\
Token Synonym&
    45.8\% & 88.3\% & 88.1\% & 63.6\% & 45.9\% \\
Paraphrase&
    42.8\% & 87.4\% & 89.4\% & 64.4\% & 36.8\% \\
Translation&
    33.7\% & 87.2\% & 88.8\% & 64.5\% & 39.6\% \\
    
    \hline
    \multicolumn{6}{c}{\textit{Model-level Evasion}} \\
    \addlinespace[2pt]
    \hline
Rank-Swap (2S) & 
    37.9\% & 63.5\% & 58.9\% & 50.0\% & 42.4\% \\
    
Rank-Swap (SS) & 
    37.9\% & 66.0\% & 71.6\% & 52.9\% & 40.6\% \\
    
Rank-Swap (SK) & 
    34.5\% & 59.9\% & 62.6\% & 55.3\% & 40.8\% \\
   
   \rowcolor{gray!25}
\textbf{Logit Laundering} & 
86.3\% {\scriptsize \textcolor{green!60!black}{(+52.5)}} &
93.7\% {\scriptsize \textcolor{green!60!black}{(+13.0)}} &
94.3\% {\scriptsize \textcolor{green!60!black}{(+7.2)}} &
94.0\% {\scriptsize \textcolor{green!60!black}{(+17.8)}} &
65.6\% {\scriptsize \textcolor{green!60!black}{(+36.2)}} \\
    \bottomrule
  \end{tabular}%
  }

  \label{tab:utility_mcq_sonnet46}
\end{table}

\begin{table}[t]
\caption{Knowledge Retention Performance on the GPT-5.4-generated MCQ set.}
  \centering
  \setlength{\tabcolsep}{4pt}
  \renewcommand{\arraystretch}{1}
  \resizebox{\columnwidth}{!}{%
  \begin{tabular}{l|c|c|c|c|c}
    \toprule
Copyright Knowledge &
LA-1B &
LA-3B &
Qw-1.5B &
Ge-2B &
SLM-1.6B\\
    \cmidrule(lr){1-6}
    \rowcolor{gray!15}
Foundation Model &
    29.8\% & 48.3\% & 56.4\% & 36.1\% & 40.9\% \\

Non Evasion &
    60.1\% & 71.9\% & 65.0\% & 73.1\% & 52.6\% \\
     \hline
    \multicolumn{6}{c}{\textit{Data-level Evasion}} \\
    \addlinespace[2pt]
    \hline
    
Char (homoglyph)  &
    58.3\% & 70.4\% & 64.6\% & 68.4\% & 50.7\% \\
Char (insertion) &
    57.0\% & 70.9\% & 65.6\% & 66.4\% & 51.8\% \\
Char (typo) &
    58.9\% & 70.3\% & 66.7\% & 58.6\% & 49.4\% \\
Token Swap&
    64.5\% & 71.8\% & 64.5\% & 63.4\% & 53.5\% \\
Token Synonym&
    53.9\% & 71.3\% & 67.8\% & 68.7\% & 56.3\% \\
Paraphrase&
    53.1\% & 70.6\% & 63.8\% & 65.3\% & 49.5\% \\
Translation&
    57.1\% & 70.5\% & 63.5\% & 64.1\% & 51.4\% \\
    
    \hline
    \multicolumn{6}{c}{\textit{Model-level Evasion}} \\
    \addlinespace[2pt]
    \hline
Rank-Swap (2S) & 
    21.3\% & 14.3\% & 22.2\% & 24.0\% & 24.6\% \\
    
Rank-Swap (SS) & 
    38.5\% & 24.7\% & 37.1\% & 37.9\% & 45.5\% \\
    
Rank-Swap (SK) & 
    25.1\% & 9.0\% & 16.2\% & 36.5\% & 49.6\% \\
   
   \rowcolor{gray!25}
\textbf{Logit Laundering} & 
59.3\% {\scriptsize \textcolor{green!60!black}{(+29.5)}} &
71.0\% {\scriptsize \textcolor{green!60!black}{(+22.7)}} &
65.8\% {\scriptsize \textcolor{green!60!black}{(+9.4)}} &
72.7\% {\scriptsize \textcolor{green!60!black}{(+36.6)}} &
52.5\% {\scriptsize \textcolor{green!60!black}{(+11.6)}} \\
    \bottomrule
  \end{tabular}%
  }

  \label{tab:utility_mcq_gpt54}
\end{table}

\subsection{Additional Experimental Results for Main Evaluation}

\noindent\textbf{Evasion Performance on STAMP-like detection variants}. Tables~\ref{tab:mainresults_stamp_details} and~\ref{tab:mainresults_stamp_details_cont} report additional results for STAMP-like auditing variants across different foundation models and evasion behaviors. Besides the standard PPL-based STAMP score used in the main text, we evaluate several scoring variants, including Min-K, Min-K++, Lowercase, and Zlib. These variants follow the same paired statistical testing framework but replace the scoring function used to measure the public/private text gap. The results show that Logit Laundering generally weakens the statistical evidence under these variants as well. Notably, some STAMP-like variants already produce non-significant results under the non-evasion setting, suggesting that their robustness as auditing signals is limited and requires further improvement.

\noindent\textbf{Knowledge Retention Performance}. 
Tables~\ref{tab:utility_mcq_sonnet46} and~\ref{tab:utility_mcq_gpt54}
disaggregate the knowledge-retention evaluation into two independently
generated MCQ sets, produced by Claude Sonnet 4.6 and GPT-5.4,
respectively. In the main text, Table~\ref{tab:utility_mcq_avg}
reports the average across these two sets to mitigate the construction
bias of any single question-generation model and thus ensure a fairer
evaluation. The appendix further shows that the same trend holds on
both sets separately: Logit Laundering remains consistently close to
the non-evasion model while substantially outperforming the original
foundation model, indicating that the copyrighted knowledge acquired
during continual pretraining is largely preserved after inference-time
reshaping. In contrast, data-level evasion methods often reduce
retention by perturbing the training texts before learning, while
Rank-Swap baselines severely damage multiple-choice performance
because answer selection is highly sensitive to top-ranked token
manipulation. Compared with the foundation model, the laundered
model still retains substantial newly acquired knowledge, indicating
that a non-compliant provider can benefit from unauthorized
copyrighted data while reducing detectability.

\noindent\textbf{General Utility Performance}. 
Table~\ref{tab:utility_general_appendix} provides detailed per-model
results for general utility. This table further breaks down the
aggregated generation-quality results reported in the main text,
including fluency (PPL), distribution quality (MAUVE), and
LLM-judge scores. Across different foundation models,
Logit Laundering generally remains close to the non-evasion model,
while the Rank-Swap baselines lead to much larger degradation in
generation quality. Inference-time logit reshaping introduces only
minor perturbations to the original decoding distribution, and
confidence-aware scaling and JS-bounded reshaping further constrain
this shift, so the impact on generation quality remains minimal. 

\begin{table}[H]
\centering
\caption{Generation quality under different evasion attacks. Lower Avg. PPL is better; higher Avg. MAUVE and judge scores are better.}
\label{tab:utility_general_appendix}
\small
\setlength{\tabcolsep}{3.5pt}
\renewcommand{\arraystretch}{0.95}
\resizebox{1\columnwidth}{!}{%
\begin{tabular}{l|l|c|c|c|c}
\toprule
\textbf{Model} & \textbf{Approach} & \textbf{Avg. PPL $\downarrow$} & \textbf{Avg. MAUVE $\uparrow$} & \textbf{GPT-5.4 as Judge$\uparrow$} & \textbf{sonnet4.6 as Judge$\uparrow$} \\
\midrule

\multirow{5}{*}{Llama-3.2-1B}
& \cellcolor{gray!15} Non Evasion      & \cellcolor{gray!15} 5.13  & \cellcolor{gray!15} 0.118 & \cellcolor{gray!15} 87.6 & \cellcolor{gray!15} 82 \\
& Rank-Swap (2S)                        & 9.30  & 0.207 & 78.3 & 80 \\
& Rank-Swap (SS)                        & 33.22 & 0.083 & 43.9 & 43 \\
& Rank-Swap (SK)                        & 11.49 & 0.238 & 72.5 & 76 \\
& \cellcolor{gray!25} Logit Laundering  & \cellcolor{gray!25} 4.29  & \cellcolor{gray!25} 0.151 & \cellcolor{gray!25} 81.8 & \cellcolor{gray!25} 79 \\

\midrule

\multirow{5}{*}{Qwen2.5-1.5B}
& \cellcolor{gray!15} Non Evasion      & \cellcolor{gray!15} 4.83  & \cellcolor{gray!15} 0.156 & \cellcolor{gray!15} 86.5 & \cellcolor{gray!15} 75 \\
& Rank-Swap (2S)                        & 9.55  & 0.175 & 77.3 & 75 \\
& Rank-Swap (SS)                        & 31.89 & 0.033 & 43.4 & 46 \\
& Rank-Swap (SK)                        & 11.60 & 0.155 & 72.2 & 74 \\
& \cellcolor{gray!25} Logit Laundering  & \cellcolor{gray!25} 5.57  & \cellcolor{gray!25} 0.151 & \cellcolor{gray!25} 82.5 & \cellcolor{gray!25} 73 \\

\midrule

\multirow{5}{*}{Llama-3.2-3B}
& \cellcolor{gray!15} Non Evasion      & \cellcolor{gray!15} 5.30  & \cellcolor{gray!15} 0.246 & \cellcolor{gray!15} 86.2 & \cellcolor{gray!15} 84 \\
& Rank-Swap (2S)                        & 10.17 & 0.231 & 77.0 & 82 \\
& Rank-Swap (SS)                        & 36.86 & 0.059 & 42.4 & 46 \\
& Rank-Swap (SK)                        & 12.66 & 0.233 & 70.4 & 77 \\
& \cellcolor{gray!25} Logit Laundering  & \cellcolor{gray!25} 5.22  & \cellcolor{gray!25} 0.194 & \cellcolor{gray!25} 82.5 & \cellcolor{gray!25} 80 \\

\midrule

\multirow{5}{*}{Gemma-2-2B}
& \cellcolor{gray!15} Non Evasion      & \cellcolor{gray!15} 7.72  & \cellcolor{gray!15} 0.151 & \cellcolor{gray!15} 88.2 & \cellcolor{gray!15} 75 \\
& Rank-Swap (2S)                        & 13.34 & 0.165 & 79.0 & 74 \\
& Rank-Swap (SS)                        & 38.77 & 0.045 & 45.3 & 47 \\
& Rank-Swap (SK)                        & 16.76 & 0.143 & 71.0 & 72 \\
& \cellcolor{gray!25} Logit Laundering  & \cellcolor{gray!25} 4.78  & \cellcolor{gray!25} 0.263 & \cellcolor{gray!25} 83.3 & \cellcolor{gray!25} 73 \\

\midrule

\multirow{5}{*}{StableLM-2-1.6B}
& \cellcolor{gray!15} Non Evasion      & \cellcolor{gray!15} 5.54  & \cellcolor{gray!15} 0.218 & \cellcolor{gray!15} 87.7 & \cellcolor{gray!15} 85 \\
& Rank-Swap (2S)                        & 9.85  & 0.222 & 77.5 & 80 \\
& Rank-Swap (SS)                        & 36.11 & 0.054 & 44.2 & 38 \\
& Rank-Swap (SK)                        & 11.74 & 0.194 & 70.7 & 78 \\
& \cellcolor{gray!25} Logit Laundering  & \cellcolor{gray!25} 8.41  & \cellcolor{gray!25} 0.518 & \cellcolor{gray!25} 87.5 & \cellcolor{gray!25} 76 \\

\bottomrule
\end{tabular}%
}
\end{table}

\suppressfloats[t]
\subsection{Additional Experimental Results for Ablations} See Table~\ref{tab:confidence_scaling_ablation_all} for the detailed ablation results across different model families.

\begin{table}[t]
  \caption{More ablation studies of confidence scaling module.}
  \centering
  \small
  \setlength{\tabcolsep}{5pt}
  \renewcommand{\arraystretch}{1}
  \resizebox{0.9\columnwidth}{!}{%
  \begin{tabular}{l l l c c}
    \toprule
    \textbf{Model} & \textbf{Variant} & \textbf{$\lambda$ sched.} & \textbf{MAUVE $\uparrow$} & \textbf{$p$-value $\uparrow$} \\
    \midrule

    \multirow{4}{*}{Llama-3.2-1B}
      & adaptive
      & $[0.17, 0.56]$
      & 0.107
      & 0.748 \\
      & w/o scaling
      & $\lambda=0.17$
      & 0.018
      & 0.044 \\
      & w/o scaling
      & $\lambda=0.365$
      & 0.016
      & 0.752 \\
      & w/o scaling
      & $\lambda=0.56$
      & 0.017
      & 0.982 \\
    \midrule

    \multirow{4}{*}{Qwen2.5-1.5B}
      & adaptive
      & $[0.14, 0.45]$
      & 0.151
      & 0.412 \\
      & w/o scaling
      & $\lambda=0.14$
      & 0.019
      & 0.069 \\
      & w/o scaling
      & $\lambda=0.295$
      & 0.017
      & 0.445 \\
      & w/o scaling
      & $\lambda=0.45$
      & 0.016
      & 0.935 \\
    \midrule

    \multirow{4}{*}{Gemma-2-2B}
      & adaptive
      & $[0.14, 0.46]$
      & 0.263
      & 0.706 \\
      & w/o scaling
      & $\lambda=0.14$
      & 0.040
      & 0.076 \\
      & w/o scaling
      & $\lambda=0.30$
      & 0.021
      & 0.710 \\
      & w/o scaling
      & $\lambda=0.46$
      & 0.016
      & 0.981 \\
    \midrule

    \multirow{4}{*}{StableLM-2-1.6B}
      & adaptive
      & $[0.12, 0.41]$
      & 0.518
      & 0.965 \\
      & w/o scaling
      & $\lambda=0.12$
      & 0.037
      & 0.194 \\
      & w/o scaling
      & $\lambda=0.265$
      & 0.044
      & 0.984 \\
      & w/o scaling
      & $\lambda=0.41$
      & 0.044
      & 1.000 \\
    \bottomrule
  \end{tabular}%
  }
  \label{tab:confidence_scaling_ablation_all}
\end{table}

\begin{table*}[p]
\caption{STAMP-like auditing performance under evasion behaviors (Part I: Min-K and Min-K++).}
  \centering
  \setlength{\tabcolsep}{5pt}
  \renewcommand{\arraystretch}{0.94}
  \resizebox{0.90\textwidth}{!}{%
  \begin{tabular}{l|cccccccccc}
    \toprule
    \multirow{1}{*}{\textbf{STAMP-Min-K}} &
    \multicolumn{2}{c}{\textbf{LLAMA-3.2-1B-Instruct}}&
    \multicolumn{2}{c}{\textbf{LLAMA-3.2-3B-Instruct}} &
    \multicolumn{2}{c}{\textbf{Qwen-2.5-1.5B-Instruct}}&
    \multicolumn{2}{c}{\textbf{Gemma-2-2B-IT}} &
    \multicolumn{2}{c}{\textbf{Stable-LM-2-1.6B-Instruct}}\\
    \cmidrule(lr){2-3}
    \cmidrule(lr){4-5}
    \cmidrule(lr){6-7}
    \cmidrule(lr){8-9}
    \cmidrule(lr){10-11}
    \textit{Metrics} &
    \textit{$p_{t}$-value $\uparrow$} & \textit{t-statistic} &
    \textit{$p_{t}$-value $\uparrow$} & \textit{t-statistic} &
    \textit{$p_{t}$-value $\uparrow$} & \textit{t-statistic} &
    \textit{$p_{t}$-value $\uparrow$} & \textit{t-statistic} &
    \textit{$p_{t}$-value $\uparrow$} & \textit{t-statistic} \\
    \midrule

\rowcolor{gray!15}
Non Evasion &
    $6.410{\times}10^{-3}$ & -3.758&
    $2.550{\times}10^{-2}$ & -2.849&
    $9.744{\times}10^{-1}$ & 10.276&
    $1.175{\times}10^{-2}$ & -3.516&
    $1.927{\times}10^{-3}$ & -4.053\\
    \hline
    \multicolumn{11}{c}{\textit{Data-level Evasion}} \\
    \addlinespace[2pt]
     \hline 
 Char (homoglyph) & 
     $5.383{\times}10^{-3}$ & -3.826& 
     $1.497{\times}10^{-1}$ & -1.611& 
     $9.760{\times}10^{-1}$ & 10.024& 
     $6.150{\times}10^{-1}$ & 0.452& 
     $8.439{\times}10^{-2}$ & -2.049\\ 
 Char (insertion) & 
     $2.860{\times}10^{-3}$ & -4.239& 
     $1.303{\times}10^{-1}$ & -1.754& 
     $9.774{\times}10^{-1}$ & 10.046& 
     $5.705{\times}10^{-1}$ & 0.268& 
     $8.421{\times}10^{-2}$ & -2.039\\ 
 Char (typo) & 
     $6.369{\times}10^{-3}$ & -3.794& 
     $1.388{\times}10^{-1}$ & -1.678& 
     $9.717{\times}10^{-1}$ & 9.640& 
     $5.966{\times}10^{-1}$ & 0.361& 
     $7.812{\times}10^{-2}$ & -2.090\\ 
 Token Swap & 
     $3.935{\times}10^{-4}$ & -4.881& 
     $9.549{\times}10^{-2}$ & -1.972& 
     $9.697{\times}10^{-1}$ & 9.668& 
     $4.883{\times}10^{-3}$ & -4.000& 
     $3.815{\times}10^{-4}$ & -4.681\\ 
 Token Synonym & 
     $7.264{\times}10^{-4}$ & -4.763& 
     $9.420{\times}10^{-2}$ & -2.026& 
     $9.654{\times}10^{-1}$ & 8.832& 
     $3.041{\times}10^{-3}$ & -4.285& 
     $4.335{\times}10^{-4}$ & -4.609\\ 
 Paraphrase & 
     $1.147{\times}10^{-3}$ & -4.720& 
     $7.226{\times}10^{-2}$ & -2.199& 
     $9.602{\times}10^{-1}$ & 7.846& 
     $6.961{\times}10^{-3}$ & -3.879& 
     $7.450{\times}10^{-4}$ & -4.456\\ 
 Translation & 
     $5.012{\times}10^{-2}$ & -2.691& 
     $2.512{\times}10^{-1}$ & -1.111& 
     $9.771{\times}10^{-1}$ & 9.986& 
     $9.683{\times}10^{-2}$ & -2.172& 
     $1.836{\times}10^{-2}$ & -3.073\\ 
     \hline
    \multicolumn{11}{c}{\textit{Model-level Evasion}} \\
    \addlinespace[2pt]
    \hline
Rank-Swap (2S) &
    $1.172{\times}10^{-2}$ & -3.415 &
    $4.363{\times}10^{-2}$ & -2.523 &
    $9.566{\times}10^{-1}$ & 7.274 &
    $1.161{\times}10^{-2}$ & -3.411 &
    $4.818{\times}10^{-3}$ & -3.622 \\

Rank-Swap (SS) &
    $7.324{\times}10^{-1}$ & 0.939 &
    $6.732{\times}10^{-1}$ & 0.704 &
    $1.000{\times}10^{0}$ & 12.735 &
    $6.632{\times}10^{-1}$ & 0.523 &
    $8.445{\times}10^{-1}$ & 1.484 \\

Rank-Swap (SK) &
    $1.729{\times}10^{-2}$ & -3.161 &
    $5.937{\times}10^{-2}$ & -2.320 &
    $9.383{\times}10^{-1}$ & 6.188 &
    $1.432{\times}10^{-2}$ & -3.165 &
    $1.065{\times}10^{-2}$ & -3.307 \\
\rowcolor{gray!25}
\textbf{Logit Laundering} &
    $6.639{\times}10^{-1}$ & 0.600&
    $8.844{\times}10^{-1}$ & 1.883&
    $9.998{\times}10^{-1}$ & 12.929&
    $5.468{\times}10^{-1}$ & 0.146&
    $9.606{\times}10^{-1}$ & 2.624\\

    \midrule

    \multirow{1}{*}{\textbf{STAMP-Min-K++}} &
    \multicolumn{2}{c}{\textbf{LLAMA-3.2-1B-Instruct}}&
    \multicolumn{2}{c}{\textbf{LLAMA-3.2-3B-Instruct}} &
    \multicolumn{2}{c}{\textbf{Qwen-2.5-1.5B-Instruct}}&
    \multicolumn{2}{c}{\textbf{Gemma-2-2B-IT}} &
    \multicolumn{2}{c}{\textbf{Stable-LM-2-1.6B-Instruct}}\\
    \cmidrule(lr){2-3}
    \cmidrule(lr){4-5}
    \cmidrule(lr){6-7}
    \cmidrule(lr){8-9}
    \cmidrule(lr){10-11}
    \textit{Metrics} &
    \textit{$p_{t}$-value $\uparrow$} & \textit{t-statistic} &
    \textit{$p_{t}$-value $\uparrow$} & \textit{t-statistic} &
    \textit{$p_{t}$-value $\uparrow$} & \textit{t-statistic} &
    \textit{$p_{t}$-value $\uparrow$} & \textit{t-statistic} &
    \textit{$p_{t}$-value $\uparrow$} & \textit{t-statistic} \\
    \midrule

\rowcolor{gray!15}
Non Evasion &
    $5.231{\times}10^{-1}$ & -4.101&
    $3.287{\times}10^{-2}$ & -3.154&
    $8.944{\times}10^{-1}$ &3.937&
    $4.966{\times}10^{-3}$ & -4.550&
    $4.947{\times}10^{-2}$ & -2.643\\
    \hline
    \multicolumn{11}{c}{\textit{Data-level Evasion}} \\
    \addlinespace[2pt]
     \hline 
 Char (homoglyph) & 
     $1.738{\times}10^{-3}$ & -4.295& 
     $1.461{\times}10^{-1}$ & -1.813& 
     $8.215{\times}10^{-1}$ & 2.735& 
     $5.824{\times}10^{-1}$ & 0.265& 
     $4.345{\times}10^{-1}$ & -0.277\\ 
 Char (insertion) & 
     $6.139{\times}10^{-4}$ & -4.997& 
     $1.134{\times}10^{-1}$ & -2.109& 
     $8.125{\times}10^{-1}$ & 2.593& 
     $6.449{\times}10^{-1}$ & 0.543& 
     $4.298{\times}10^{-1}$ & -0.293\\ 
 Char (typo) & 
     $3.700{\times}10^{-3}$ & -4.128& 
     $1.487{\times}10^{-1}$ & -1.813& 
     $6.137{\times}10^{-1}$ & 0.557& 
     $6.664{\times}10^{-1}$ & 0.616& 
     $3.962{\times}10^{-1}$ & -0.451\\ 
 Token Swap & 
     $1.780{\times}10^{-3}$ & -5.546& 
     $9.947{\times}10^{-2}$ & -2.118& 
     $9.141{\times}10^{-1}$ & 4.607& 
     $2.183{\times}10^{-3}$ & -4.933& 
     $3.015{\times}10^{-2}$ & -3.116\\ 
 Token Synonym & 
     $8.908{\times}10^{-4}$ & -5.475& 
     $8.341{\times}10^{-2}$ & -2.278& 
     $4.483{\times}10^{-1}$ & -0.608& 
     $4.681{\times}10^{-4}$ & -5.599& 
     $2.209{\times}10^{-2}$ & -3.170\\ 
 Paraphrase & 
     $6.097{\times}10^{-4}$ & -5.341& 
     $7.293{\times}10^{-2}$ & -2.628& 
     $1.855{\times}10^{-2}$ & -4.630& 
     $1.031{\times}10^{-2}$ & -4.047& 
     $3.935{\times}10^{-2}$ & -2.884\\ 
 Translation & 
     $2.173{\times}10^{-2}$ & -3.294& 
     $2.631{\times}10^{-1}$ & -1.122& 
     $3.993{\times}10^{-1}$ & -0.812& 
     $5.189{\times}10^{-2}$ & -2.659& 
     $1.605{\times}10^{-1}$ & -1.519\\ 
     \hline 
    \multicolumn{11}{c}{\textit{Model-level Evasion}} \\
    \addlinespace[2pt]
    \hline
Rank-Swap (2S) &
    $6.243{\times}10^{-1}$ & 0.439 &
    $6.214{\times}10^{-1}$ & 0.363 &
    $2.998{\times}10^{-2}$ & -3.871 &
    $4.730{\times}10^{-1}$ & -0.126 &
    $6.336{\times}10^{-1}$ & 0.449 \\

Rank-Swap (SS) &
    $8.211{\times}10^{-1}$ & 1.272 &
    $7.116{\times}10^{-1}$ & 0.884 &
    $8.002{\times}10^{-2}$ & -2.887 &
    $6.439{\times}10^{-1}$ & 0.511 &
    $7.221{\times}10^{-1}$ & 0.859 \\

Rank-Swap (SK) &
    $6.492{\times}10^{-1}$ & 0.528 &
    $6.299{\times}10^{-1}$ & 0.394 &
    $3.493{\times}10^{-2}$ & -3.777 &
    $4.961{\times}10^{-1}$ & -0.048 &
    $6.626{\times}10^{-1}$ & 0.544 \\

\rowcolor{gray!25}
\textbf{Logit Laundering} &
    $1.221{\times}10^{-1}$ & -1.838 &
    $7.555{\times}10^{-1}$ & 0.971 &
    $1.000{\times}10^{0}$ & 15.024&
    $3.744{\times}10^{-1}$ & -0.506&
    $9.853{\times}10^{-1}$ & 3.434\\
    \bottomrule
  \end{tabular}%
  }

  \label{tab:mainresults_stamp_details}
\end{table*}

\begin{table*}[p]
\caption{Comparison of STAMP-like auditing performance under evasion behaviors (Part II: Lowercase and Zlib).}
  \centering
  \setlength{\tabcolsep}{5pt}
  \renewcommand{\arraystretch}{0.94}
  \resizebox{0.90\textwidth}{!}{%
  \begin{tabular}{l|cccccccccc}
    \toprule
    \multirow{1}{*}{\textbf{STAMP-Lowercase}} &
    \multicolumn{2}{c}{\textbf{LLAMA-3.2-1B-Instruct}}&
    \multicolumn{2}{c}{\textbf{LLAMA-3.2-3B-Instruct}} &
    \multicolumn{2}{c}{\textbf{Qwen-2.5-1.5B-Instruct}}&
    \multicolumn{2}{c}{\textbf{Gemma-2-2B-IT}} &
    \multicolumn{2}{c}{\textbf{Stable-LM-2-1.6B-Instruct}}\\
    \cmidrule(lr){2-3}
    \cmidrule(lr){4-5}
    \cmidrule(lr){6-7}
    \cmidrule(lr){8-9}
    \cmidrule(lr){10-11}
    \textit{Metrics} &
    \textit{$p_{t}$-value $\uparrow$} & \textit{t-statistic} &
    \textit{$p_{t}$-value $\uparrow$} & \textit{t-statistic} &
    \textit{$p_{t}$-value $\uparrow$} & \textit{t-statistic} &
    \textit{$p_{t}$-value $\uparrow$} & \textit{t-statistic} &
    \textit{$p_{t}$-value $\uparrow$} & \textit{t-statistic} \\
    \midrule

\rowcolor{gray!15}
Non Evasion &
    $1.434{\times}10^{-1}$ & -2.843&
    $2.377{\times}10^{-1}$ & -1.521&
    $3.487{\times}10^{-2}$ & -4.628&    
$2.956{\times}10^{-2}$ & -4.559&
    $8.572{\times}10^{-3}$ & -3.894\\
    \hline
    \multicolumn{11}{c}{\textit{Data-level Evasion}} \\
    \addlinespace[2pt]
     \hline 
 Char (homoglyph) & 
     $1.740{\times}10^{-1}$ & -2.371& 
     $4.776{\times}10^{-1}$ & -0.110& 
     $3.850{\times}10^{-2}$ & -4.456& 
     $3.505{\times}10^{-1}$ & -0.588& 
     $6.259{\times}10^{-1}$ & 0.526\\ 
 Char (insertion) & 
     $1.405{\times}10^{-1}$ & -2.858& 
     $4.766{\times}10^{-1}$ & -0.120& 
     $3.585{\times}10^{-2}$ & -4.676& 
     $3.098{\times}10^{-1}$ & -0.703& 
     $5.895{\times}10^{-1}$ & 0.329\\ 
 Char (typo) & 
     $1.537{\times}10^{-1}$ & -2.519& 
     $5.153{\times}10^{-1}$ & 0.096& 
     $3.529{\times}10^{-2}$ & -5.368& 
     $2.557{\times}10^{-1}$ & -0.975& 
     $6.265{\times}10^{-1}$ & 0.520\\ 
 Token Swap & 
     $1.010{\times}10^{-1}$ & -3.448& 
     $4.500{\times}10^{-1}$ & -0.272& 
     $2.950{\times}10^{-2}$ & -5.760& 
     $1.138{\times}10^{-3}$ & -4.624& 
     $8.037{\times}10^{-3}$ & -4.222\\ 
 Token Synonym & 
     $9.573{\times}10^{-2}$ & -3.733& 
     $3.990{\times}10^{-1}$ & -0.478& 
     $2.565{\times}10^{-2}$ & -6.312& 
     $3.370{\times}10^{-4}$ & -5.087& 
     $6.685{\times}10^{-3}$ & -4.363\\ 
 Paraphrase & 
     $8.578{\times}10^{-2}$ & -3.959& 
     $3.735{\times}10^{-1}$ & -0.606& 
     $8.311{\times}10^{-6}$ & -9.738& 
     $2.330{\times}10^{-2}$ & -3.380& 
     $6.691{\times}10^{-3}$ & -4.302\\ 
 Translation & 
     $1.766{\times}10^{-1}$ & -2.281& 
     $5.395{\times}10^{-1}$ & 0.224& 
     $1.265{\times}10^{-1}$ & -3.019& 
     $2.105{\times}10^{-1}$ & -1.307& 
     $4.423{\times}10^{-2}$ & -2.982\\ 
     \hline 
    \multicolumn{11}{c}{\textit{Model-level Evasion}} \\
    \addlinespace[2pt]
    \hline
Rank-Swap (2S) &
    $2.601{\times}10^{-1}$ & -1.069 &
    $3.890{\times}10^{-1}$ & -0.465 &
    $1.130{\times}10^{-1}$ & -1.950 &
    $2.669{\times}10^{-1}$ & -1.045 &
    $1.601{\times}10^{-1}$ & -1.540 \\

Rank-Swap (SS) &
    $4.978{\times}10^{-1}$ & -0.063 &
    $5.148{\times}10^{-1}$ & 0.022 &
    $5.713{\times}10^{-1}$ & 0.289 &
    $4.700{\times}10^{-1}$ & -0.079 &
    $4.987{\times}10^{-1}$ & 0.021 \\

Rank-Swap (SK) &
    $3.151{\times}10^{-1}$ & -0.729 &
    $4.106{\times}10^{-1}$ & -0.362 &
    $1.837{\times}10^{-1}$ & -1.327 &
    $3.596{\times}10^{-1}$ & -0.644 &
    $2.448{\times}10^{-1}$ & -1.068 \\

\rowcolor{gray!25}
\textbf{Logit Laundering} &
    $8.886{\times}10^{-1}$ & 2.093&
    $9.859{\times}10^{-1}$ & 3.447&
    $9.944{\times}10^{-1}$ & 6.091&
    $8.721{\times}10^{-1}$ & 1.892&
    $9.635{\times}10^{-1}$ & 3.415\\

    \midrule

    \multirow{1}{*}{\textbf{STAMP-Zlib}} &
    \multicolumn{2}{c}{\textbf{LLAMA-3.2-1B-Instruct}}&
    \multicolumn{2}{c}{\textbf{LLAMA-3.2-3B-Instruct}} &
    \multicolumn{2}{c}{\textbf{Qwen-2.5-1.5B-Instruct}}&
    \multicolumn{2}{c}{\textbf{Gemma-2-2B-IT}} &
    \multicolumn{2}{c}{\textbf{Stable-LM-2-1.6B-Instruct}}\\
    \cmidrule(lr){2-3}
    \cmidrule(lr){4-5}
    \cmidrule(lr){6-7}
    \cmidrule(lr){8-9}
    \cmidrule(lr){10-11}
    \textit{Metrics} &
    \textit{$p_{t}$-value $\uparrow$} & \textit{t-statistic} &
    \textit{$p_{t}$-value $\uparrow$} & \textit{t-statistic} &
    \textit{$p_{t}$-value $\uparrow$} & \textit{t-statistic} &
    \textit{$p_{t}$-value $\uparrow$} & \textit{t-statistic} &
    \textit{$p_{t}$-value $\uparrow$} & \textit{t-statistic} \\
    \midrule

\rowcolor{gray!15}
Non Evasion &
    $1.499{\times}10^{-2}$ & -3.012&
    $2.570{\times}10^{-2}$ & -2.744&
    $3.492{\times}10^{-1}$ & -0.790&
    $2.155{\times}10^{-2}$ & -3.114&
    $7.164{\times}10^{-5}$ & -5.881\\
    \hline
    \multicolumn{11}{c}{\textit{Data-level Evasion}} \\
    \addlinespace[2pt]
     \hline 
 Char (homoglyph) & 
     $1.661{\times}10^{-2}$ & -2.978& 
     $4.896{\times}10^{-2}$ & -2.440& 
     $3.182{\times}10^{-1}$ & -0.965& 
     $2.980{\times}10^{-1}$ & -0.797& 
     $9.245{\times}10^{-4}$ & -4.821\\ 
 Char (insertion) & 
     $1.293{\times}10^{-2}$ & -3.087& 
     $4.672{\times}10^{-2}$ & -2.466& 
     $3.225{\times}10^{-1}$ & -0.946& 
     $2.665{\times}10^{-1}$ & -0.951& 
     $9.361{\times}10^{-4}$ & -4.812\\ 
 Char (typo) & 
     $1.578{\times}10^{-2}$ & -2.999& 
     $4.800{\times}10^{-2}$ & -2.453& 
     $3.130{\times}10^{-1}$ & -0.996& 
     $2.715{\times}10^{-1}$ & -0.928& 
     $8.812{\times}10^{-4}$ & -4.831\\ 
 Token Swap & 
     $9.385{\times}10^{-3}$ & -3.208& 
     $4.469{\times}10^{-2}$ & -2.498& 
     $3.143{\times}10^{-1}$ & -0.993& 
     $1.810{\times}10^{-2}$ & -3.237& 
     $4.976{\times}10^{-5}$ & -5.979\\ 
 Token Synonym & 
     $9.331{\times}10^{-3}$ & -3.219& 
     $4.020{\times}10^{-2}$ & -2.546& 
     $2.847{\times}10^{-1}$ & -1.178& 
     $9.697{\times}10^{-3}$ & -3.584& 
     $3.131{\times}10^{-5}$ & -6.102\\ 
 Paraphrase & 
     $9.759{\times}10^{-3}$ & -3.167& 
     $4.189{\times}10^{-2}$ & -2.514& 
     $2.405{\times}10^{-1}$ & -1.428& 
     $1.398{\times}10^{-2}$ & -3.379& 
     $4.098{\times}10^{-5}$ & -6.074\\ 
 Translation & 
     $3.611{\times}10^{-2}$ & -2.617& 
     $6.867{\times}10^{-2}$ & -2.268& 
     $4.138{\times}10^{-1}$ & -0.420& 
     $9.277{\times}10^{-2}$ & -2.051& 
     $1.554{\times}10^{-4}$ & -5.553\\ 
     \hline 
    \multicolumn{11}{c}{\textit{Model-level Evasion}} \\
    \addlinespace[2pt]
    \hline
Rank-Swap (2S) &
    $1.923{\times}10^{-2}$ & -2.853 &
    $3.181{\times}10^{-2}$ & -2.614 &
    $3.883{\times}10^{-1}$ & -0.595 &
    $2.904{\times}10^{-2}$ & -2.944 &
    $1.266{\times}10^{-4}$ & -5.711 \\

Rank-Swap (SS) &
    $6.482{\times}10^{-2}$ & -2.302 &
    $8.014{\times}10^{-2}$ & -2.193 &
    $7.614{\times}10^{-1}$ & 2.147 &
    $1.286{\times}10^{-1}$ & -1.647 &
    $6.475{\times}10^{-3}$ & -4.451 \\

Rank-Swap (SK) &
    $2.436{\times}10^{-2}$ & -2.761 &
    $3.763{\times}10^{-2}$ & -2.543 &
    $4.401{\times}10^{-1}$ & -0.315 &
    $3.700{\times}10^{-2}$ & -2.767 &
    $1.702{\times}10^{-4}$ & -5.604 \\

\rowcolor{gray!25}
\textbf{Logit Laundering} &
  $1.680{\times}10^{-1}$ & -1.655&
    $2.013{\times}10^{-1}$ & -1.452&
    $7.636{\times}10^{-1}$ & 1.943&
    $3.159{\times}10^{-1}$ & -0.692&
    $2.155{\times}10^{-2}$ & -3.271 \\

    \bottomrule
  \end{tabular}%
  }

  \label{tab:mainresults_stamp_details_cont}
\end{table*}

\end{document}